\documentclass[
 reprint,
notitlepage,
superscriptaddress,
 amsmath,
 amssymb,
 aps,
floatfix,
onecolumn,
]{revtex4-2}

\usepackage{graphicx}
\usepackage{dcolumn}
\usepackage{color}
\usepackage{xcolor}
\usepackage{subfigure}
\usepackage{amsmath}
\usepackage{amssymb} 
\usepackage{color}
\usepackage{multirow}
\usepackage[normalem]{ulem}
\usepackage{bm}
\usepackage{array}

\begin{document}

\title{The effects of interparticle cohesion on the collapse of granular columns}

\author{Ram Sudhir Sharma}
\email{ramsharma@ucsb.edu}
\affiliation{Department of Mechanical Engineering, University of California, Santa Barbara, CA 93106, USA}
\author{Wladimir Sarlin}
\affiliation{Laboratoire d'Hydrodynamique, CNRS, École polytechnique, Institut Polytechnique de Paris, 91120 Palaiseau, France}
\affiliation{Universit\'e Paris-Saclay, CNRS, Laboratoire FAST, 91405 Orsay, France}
\author{Langqi Xing}
\affiliation{Department of Mechanical Engineering, University of California, Santa Barbara, CA 93106, USA}
\author{Cyprien Morize}
\affiliation{Universit\'e Paris-Saclay, CNRS, Laboratoire FAST, 91405 Orsay, France}
\author{Philippe Gondret}
\affiliation{Universit\'e Paris-Saclay, CNRS, Laboratoire FAST, 91405 Orsay, France}
\author{Alban Sauret}
\email{asauret@ucsb.edu}
\affiliation{Department of Mechanical Engineering, University of California, Santa Barbara, CA 93106, USA}

\date{\today}

\begin{abstract}
The presence of interparticle cohesion can drastically change the behavior of granular materials. For instance, powders are challenging to handle, and one can make a sandcastle using wet grains. In this study, we report experimental results for columns of model cohesive grains collapsing under their own weight in air and spreading on a rough horizontal surface. The effects of two different sources of interparticle cohesion on two collapse geometries are compared and rationalized in a common framework. Grains are made cohesive by adding a small amount of water, such that they are in the pendular state, or by applying a polymer coating. The effects of cohesion are reported for a cylindrical column that spreads unconfined axisymmetrically and a confined rectangular column that flows in a single direction. A dimensionless number, comparing macroscopic cohesive strength to particle weight, is shown to capture the effects of cohesion on the final morphology. To this end, a characterization of the cohesive strength of the granular materials is obtained, independent of the physical source of cohesion at the particle scale. Such a framework allows for a common description of cohesive granular materials with different sources of cohesion.
\end{abstract}

\maketitle


\section{Introduction}
\label{sec:sec1_Introduction}

Cohesion, or adhesive forces between particles, is ubiquitous in natural systems and observed in many industrial settings. In industry, the presence of cohesion can lead to problems in processing, such as reduced flowability and caking of powders \cite{2007_schulze}. Cohesion also results in the agglomeration of particles \cite{2018_raux}, which can be useful for some processes, such as powder coating \cite{2007_schulze}. Adhesive forces between particles can be caused by various physical mechanisms. In dry granular materials, inter-particle adhesion can be induced by electrostatic forces dependent on the surface properties of the particles or by Van der Waals forces \cite{2007_schulze}. The presence of moisture can result in capillary bridges between particles, where surface tension causes particles to attract \cite{herminghaus2005dynamics,2006_mitarai, 2007_schulze}. These forces depend on the distance between particles. For an arbitrary assembly of cohesive particles, it is not always straightforward to discern what the primary source of cohesion may be. For instance, powders may be sticky due to both Van der Waals forces and capillary bridges \cite{2009_royer}.
\smallskip

In natural systems, such as soil, several sources of inter-particle adhesion may be at play simultaneously. Besides Van der Waals and capillary forces, the presence of bacterial biofilms \cite{2008_garrett} may cause adhesion. The presence of a wide variety of particle shapes can also modify bulk yield properties \cite{2011_brown}. When cohesive soils are destabilized, for instance, as a result of rapid erosion due to drop impacts during heavy rainfalls \cite{2022_cheng,kostynick2022rheology}, large-scale mudflows, debris flows, or landslides can be triggered, carrying massive amounts of material over large distances. Many features of these complex geophysical flows are recovered when modeling them experimentally with granular media \cite{2006_lajeunesse,2009_staron,2019_jerolmack}. This is one reason why considerable progress has been made in the last two decades towards a description of granular media, especially by employing continuum approaches \cite{2011_lagree,2015_ionescu} like the $\mu(I)$ rheology for dry cohesionless granular flows \cite{2006_jop}. However, the possible extension of such approaches to cohesive granular materials remains elusive.
\smallskip

To obtain a better understanding of landslides at the laboratory scale, experiments of gravity-driven collapses of columns of cohesionless grains were first performed by Lajeunesse \textit{et al.} \cite{2004_lajeunesse}, Lube \textit{et al.} \cite{2004_lube}, and Balmforth and Kerswell \cite{2005_balmforth}. Scaling laws based on these early studies were successfully applied, for instance, to describe the runout of large landslides on Mars \cite{2006_lajeunesse}. Hence, for almost two decades, {numerous} studies have been dedicated to the collapse of dry cohesionless granular columns \cite{2005_lajeunesse,2005_lube,2005_staron,2008_lacaze,2011_lagree,2015_ionescu, 2021a_man}. In this model experiment, a granular column of initial height $H_0$ and radius $R_0$ (for a cylindrical column) or width $L_0$ (for a rectangular column) is {suddenly} released and collapses under the effect of gravity. The cylindrical column relaxes unconfined on a flat surface, while the rectangular column spreads confined within a channel. When the final heap stabilizes, one can characterize the resulting deposit by two typical lengths \cite{2004_lube}: The final maximum height, $H_\infty$, and the runout distance $\Delta R_\infty$ (axisymmetric) or $\Delta L_\infty$ (quasi-two-dimensional). The runout is defined as the {total} distance traveled in the spreading direction, \textit{i.e.}, $\Delta R_{\infty} = R_{\infty} - R_0$. The initial aspect ratio of the column, defined as $a=H_0/R_0$ (respectively, $a=H_0/L_0$ for the rectangular geometry), governs the final morphology of the deposit in both geometries by empirical piece-wise power laws \cite{2004_lube, 2005_lajeunesse}.
\smallskip

In the axisymmetric case, the final height rescaled by the initial radius, $H_\infty/R_0$ equals $a$ for small {aspect ratios} $a$ and is {roughly constant} for larger $a$. For the quasi-2D geometry, the rescaled final height $H_{\infty}/L_0$ was similarly found to be equal to $a$ for small $a$, but scale with $a^{1/3}$ for large initial aspect ratio. In the small $a$ cases{, and} for both geometries, the {column only collapses on the edges, such that} $H_{\infty} = H_0$. The rescaled runout in the axisymmetric case, $\Delta R_{\infty}/R_0$, was found to scale with $a$ for small $a$, and with $a^{1/2}$ for large $a$, whereas for the 2D case ${\Delta} L_{\infty}/L_0$ was also found to scale with $a$ for small $a$, and to evolve with $a^{2/3}$ for large initial $a$. The critical aspect ratios that mark the transitions between the collapse regimes were also found to differ between the two geometries \cite{2005_lajeunesse}. The values for the scaling {coefficients} as well as the critical aspect ratios, and thus the power laws {themselves}, are dependent on the material and frictional properties of the grains \cite{2005_balmforth,2021a_man}, and may also be impacted by the presence of {some} finite-size effects \cite{2019_cabrera}. Many experimental studies \cite{2005_balmforth,2006_meriaux,2008_lacaze,2013_degaetano,2014_warnett,2019_cabrera,2021a_man,2021c_sarlin} and numerical simulations \cite{2005_staron,2005_zenit,2006_larrieu,2007_staron,2009_lacaze,2011_lagree,2012_tapia-mcclung,2015_ionescu} have since retrieved similar scaling behavior for cohesionless collapses. However, various granular flows involved in geophysical or industrial processes feature additional physical richness, which explains why the behavior of the granular collapse was found to be significantly altered in cases where the medium is fluidized \cite{2011_roche}, polydisperse \cite{2013_degaetano,2022_martinez}, or if adhesive forces exist between the particles \cite{2013_artoni,2018_santomaso,2021_li, 2022_li, 2023_gans}. In particular, cohesion is known to alter the rheological and mechanical properties of granular materials \cite{2020_mandal, 1992_nedderman, 2007_moller}. In the context of granular collapse, the effects of a small amount of water on the stability of a granular column has been of particular interest \cite{2005_schiffer, 2005_nowak, 2008_scheel,2012_pakpour}.
\smallskip

Modeling and controlling cohesion experimentally remains challenging. Following the common experience of building sandcastles, one method consists of mixing a certain amount of liquid with the grains to constitute an unsaturated wet granular medium \cite{2004_kohonen,2006_mitarai,saingier2017accretion}. Different mechanical properties are bestowed to the material depending on the fraction of liquid added to it $W_{\%}$, which determines whether it belongs to the pendular, funicular, capillary, or slurry states (for increasing amounts of added fluid, respectively) \cite{2006_mitarai}. In the pendular regime, the added liquid is contained within capillary bridges between particles, which holds them together and makes the material cohesive. This approach has been used to investigate the collapse of rectangular cohesive columns in both channelized \cite{2013_artoni,2018_santomaso,2021_li} and unchannelized collapse geometries \cite{2022_li}. Artoni \textit{et al.} \cite{2013_artoni} discussed the influence of cohesion on the final morphology of the deposit in terms of two parameters: a dimensionless Bond number $\mathrm{Bo}$, which compares the effects of gravity and surface tension, and the relative mass of liquid added to the granular material, $W_{\%}$. Santomaso \textit{et al.} \cite{2018_santomaso} compared the collapse of columns made of spherical beads {to those made of} natural polydisperse sand for various water contents and evidenced significant differences between them. More recently, Li \textit{et al.} \cite{2021_li,2022_li} varied the aspect ratio $a$, the particle diameter $d$, and the added water amount $W_{\%}$ for the channelized and unchannelized collapse of rectangular columns. These authors provide a phase diagram of the different collapse regimes: The continuous collapse (in the absence or the presence of a weak cohesion), the block collapse (cohesive case), and a stable region where no collapse occurs. For the unchannelized collapse of a rectangular column, Li \textit{et al.} \cite{2022_li} propose scaling laws for the final morphology of the deposit, which capture the effects of $W_{\%}$ and $\mathrm{Bo}$ and connect with the cohesionless case.
\smallskip

Another recent approach developed to {produce} cohesive grains consists of coating glass beads with a thin layer of polymers. Cohesion is controlled by tuning the coating thickness \cite{2016_hemmerle,2020_gans}. For instance, this cohesion-controlled granular material (CCGM) has been used {as a model material} to study the effects of cohesion on erosion induced by turbulent air jets \cite{2022_sharma} and {on} the discharge of silos \cite{gans2021effect}. Compared to capillary bridges, this source of cohesion has the advantage of avoiding drainage and evaporation of the menisci, which could play a significant role over time. Gans \textit{et al.} \cite{2023_gans} have used such cohesive grains to study the stability and collapse of a confined rectangular column in the quasi-2D geometry. In this work, the authors combine experiments with {numerical} simulations {based on} a continuum model to discuss the stability and failure of quasi-bidimensional columns of cohesive grains. While not explicitly compared, the effects of cohesion provided in this study appear, at first glance, to be qualitatively similar to the morphology of deposits for wet grains in the pendular state, where $\Delta R_{\infty}$ is reduced while $H_{\infty}$ becomes larger when the {interparticle} cohesive force increases. Other kinds of apparent cohesion have been used in geometries similar to collapsing columns. Sarate \textit{et al.} \cite{2022_sarate}, for instance, studied the deposition of flexible granular chains, showing where such systems display stability rather than spreading. Several studies have also demonstrated how the physical entanglement of dry particles can also display cohesion-like stability using a similar collapse geometry \cite{2016_gravish,2008_zhao,2020_weiner}. Overall, various experimental studies have {investigated} the influence of different sources of cohesion, but a common framework rationalizing the macroscopic observations remains elusive.
\smallskip

Numerical simulations have also been used to probe the role of cohesion on 2D collapses, using contact models of cohesion \cite{2020_abramian, 2021_abramian, 2015_langlois, 2022_zhu}. Abramian \textit{et al.} \cite{2020_abramian} compared discrete simulations to a continuum model with cohesion and provided a framework to connect micro- and macro-scale effects for a rectangular channelized geometry. Further{more}, they also examined the correlation between roughness {on the surface of the deposit} and cohesive forces {between the grains} \cite{2021_abramian}. Other numerical simulations, for instance, by Langlois \textit{et al.} \cite{2015_langlois} for dry brittle cohesive columns and by Zhu \textit{et al.} \cite{2022_zhu} for submerged ones, investigated the collapse process in the channelized rectangular configurations. These studies provided, among other insights, details into the internal structure of the collapsed pile and the trajectories and size of particle clusters.
\smallskip 

Despite the recent research interest in the collapse of cohesive granular columns, several points remain elusive. In particular, it is unclear whether a common framework can capture the macroscopic effects of cohesion on the collapses regardless of the source of cohesion at the particle scale. The development of such a macroscopic framework is of great significance since cohesive granular assemblies encountered in natural {environments} can typically have several sources of cohesion acting simultaneously. In addition, most studies on cohesive collapses have focused on quasi-2D rectangular columns in a confined channel, while natural situations typically occur in a tri-dimensional context. In this regard, it is relevant to better understand unconfined situations and investigate whether the behavior observed in two-dimensional collapses applies to more realistic geometries.
\smallskip 

In the present study, we report {experiments} on collapsing granular columns for various ranges of cohesion, grain size, and column geometries. We compare the effects of cohesion on the final morphology of deposits across the 3D axisymmetric and 2D channelized geometries, with a particular focus on the former. The initial aspect ratio $a$ of the column and the cohesive forces between the grains are systematically varied. Two different sources of cohesion are considered: Wet granular material in the pendular state, where water is the wetting liquid, and a polymer coating applied to glass beads to obtain a CCGM. In Sec. \ref{sec:sec2_Methods}, we detail the experimental apparatus and methods. We then characterize the macroscopic cohesion from the two sources into a common framework in Sec. \ref{sec:sec3_Cohesion}. To this end, a dimensionless number {$\mathrm{Co}$}, comparing the bulk cohesion to the particle weight, is constructed by performing yield strength measurements from independent experiments {using} a shear cell. Results of collapsing axisymmetric and rectangular columns and the morphology of the final deposits for both geometries are provided in Sec. \ref{sec:sec4_Results}. Finally, in Sec. \ref{sec:sec5_Discussion}, the effects of cohesion on the piece-wise power laws are rationalized using the macroscopic cohesive number $\mathrm{Co}$. An alternative interpretation of this dimensionless number as the ratio of the cohesive length to the grain diameter is also discussed, which enables us to connect the microscopic and macroscopic cohesion across the two {collapse} geometries.
\smallskip


\begin{figure}
\centering
\includegraphics[width = 0.85\textwidth]{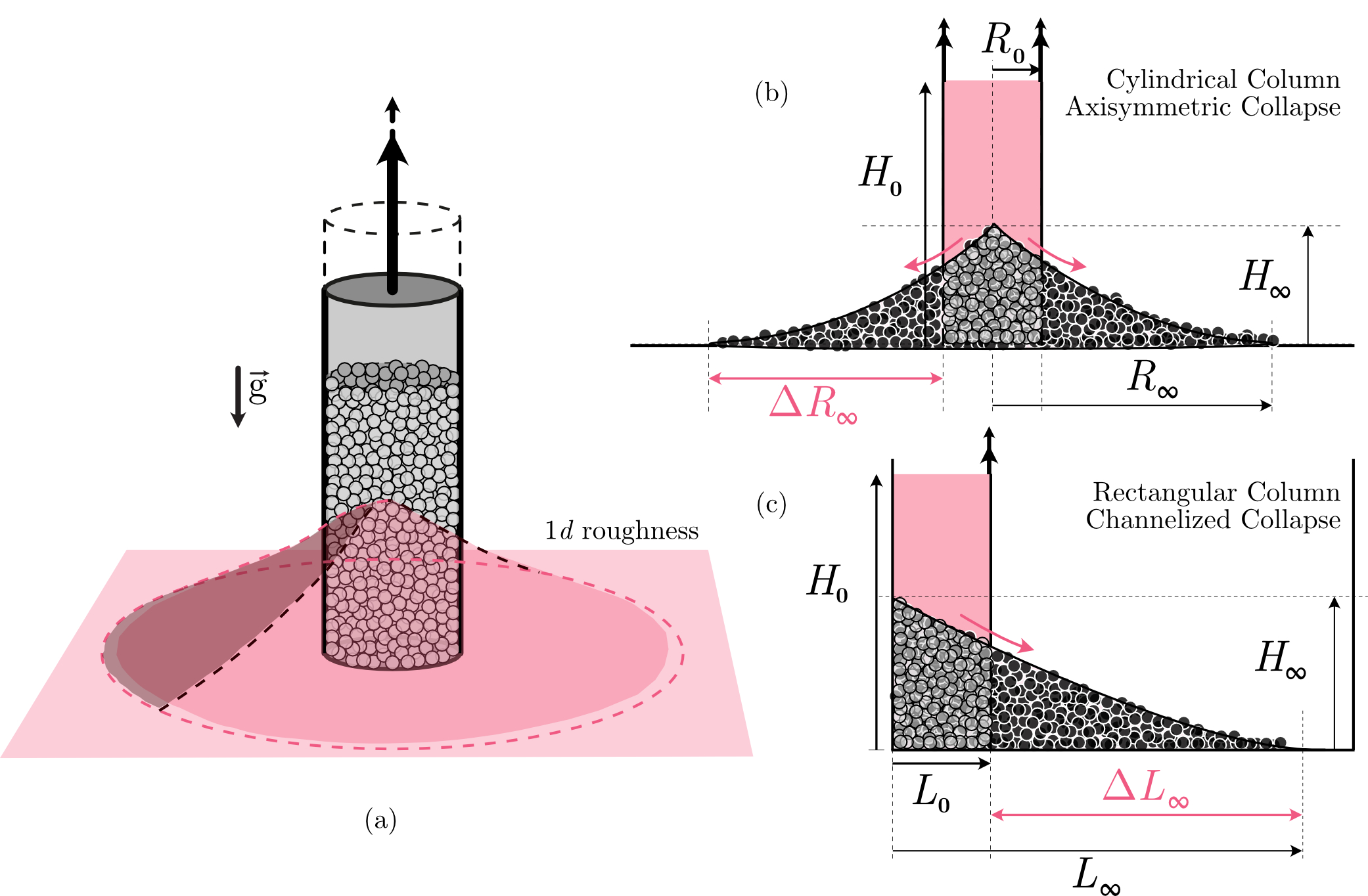}
\caption{(a) Schematic of the 3D axisymmetric experimental apparatus. A hollow cylinder is first filled with a granular material, then the cylindrical shell is rapidly lifted, which allows the grains to spread axisymmetrically on a horizontal surface under their own weight {,as illustrated by the dashed lines}. (b) Side view of the 3D axisymmetric setup showing the initial dimensions of the cylindrical column $H_0$ and $R_0$, and {the final height $H_{\infty}$ and runout $R_{\infty}$ of the relaxed pile}. (c) Side view of the 2D channelized setup showing the initial dimensions of the rectangular column $H_0$ and $L_0$, and the final deposit {of final height $H_{\infty}$ and runout distance $\Delta L_{\infty} = L_{\infty} - L_0$}. The horizontal {bottom} surfaces for both cases are coated with one layer of {glued} grains.}
\label{fig:Fig1_Schematic}
\end{figure}

\section{Experimental Methods}
\label{sec:sec2_Methods}


\subsection{Experiment Setup}
\label{subsec:sec2_1_Methods}

For the unconfined collapse of an axisymmetric column, a hollow poly-methyl methacrylate (PMMA) cylindrical shell is initially filled with grains of diameter $d$, as illustrated in Fig. \ref{fig:Fig1_Schematic}(a). We use two different cylinders with internal radii, $R_0$ = 4.2 cm and 2.5 cm, respectively. By controlling the amount of grains filling each cylinder, we vary the initial height of the column $H_0$. {An identical loading procedure is used for both cohesionless and cohesive grains, even if the presence of cohesion leads to a slightly smaller volume fraction \cite{2020_gans}.} Grains are poured in {using a funnel} and gently flattened to resemble a cylindrical column without introducing any noticeable compaction. We use the cylinder of internal radius $R_0$ = 4.2 cm to make columns of aspect ratios $a \equiv H_0 / R_0 \in [0.25,\, {6}]$ and the cylinder of radius $R_0$ = 2.5  cm to make columns with $a \in [{3}, 10]$. This approach allows us to investigate a range of aspect ratios $a \in [0.25, 10]$, {which is similar to most previous studies on the topic \cite{2004_lube,2004_lajeunesse,2005_lube,2005_lajeunesse,2005_staron,2007_thompson,2019_cabrera,2021a_man}}. Even for the largest grains used and the narrowest cylinder, $R_0 / d \approx 25$, suggesting that the confinement effects remain {negligible} \cite{2021b_man}. {For each batch of grains (cohesionless or cohesive), a trial was repeated using both the narrowest ($R_0 = 2.5$ cm) and the widest ($R_0 = 4.2$ cm) cylinders for $a$ between 3 and 6, to quantitatively ensure that no significant finite size effects were at play.}  The cylinder is connected to a pneumatic system, allowing for its rapid vertical translation over its full length at a velocity of order $2$  m.s$^{-1}$. At this velocity, the effect of the release does not influence the collapse dynamics \cite{2021c_sarlin}. The whole system is placed on a rough surface, made by gluing $\sim$ one layer of grains of $d=0.7$ mm on a large base. When the cylinder is lifted, the initial column follows an unconstrained collapse and spreads radially on the rough surface. Once the grains have come to rest, an axisymmetric deposit is observed, as illustrated by the dashed lines in Fig. \ref{fig:Fig1_Schematic}(a). A camera images the collapse from the side [Fig. \ref{fig:Fig1_Schematic}(b)]. For this geometry, the final height of the relaxed pile is denoted $H_{\infty}$ and is located at the center, and the radial runout is denoted $\Delta R_{\infty}$.
\smallskip

A PMMA tank is used for the channelized collapse of an initially rectangular column. A removable gate is used on one side to enclose a rectangular section of grains for the entire spanwise width of the tank (20 cm), and the bottom surface is coated with a layer of grains. A side view as imaged by a camera is schematized in Fig. \ref{fig:Fig1_Schematic}(c). We then vary the two dimensions of the initial column, height $H_0$ and length $L_0$, defining an aspect ratio $a \equiv H_0 / L_0$ similar to the unconfined cylindrical geometry. For the experiments considered here, we use three initial lengths $L_0$ = 1.6, 5, and 10 cm. We use $L_0$ = 10 cm to make columns of $a \equiv H_0 / L_0 \in [0.25, \, 1.5]$; $L_0$ = 5 cm to make $a \in [1.5, \, 4]$; and $L_0$ = {1.6} cm to make $a \in [4, \, 12]$. We repeat some aspect ratios for different $L_0$ to confirm that the confinement effects {remain} small. Altogether, the range $a \in [0.25, \, 12]$ comprises a similar range to the {axisymmetric} geometry. The relaxing column is channeled by the width of the confining tank. The final height of the deposit is maximum at the back wall. The extensional runout is also measured similarly to the axisymmetric case, \textit{i.e.}, $\Delta L_{\infty} = L_{\infty} - L_0$.
\smallskip

All columns are composed of glass spheres (Potters, Inc.) with density $\rho$ = 2.5 g.cm$^{-3}$, {which can be made cohesive as discussed in the next section}. In this study, we use four different grain diameters: $d$= 1.1 mm ($\bigcirc$),  $d$=0.7 mm ($\Diamond$), $d$ = 0.5 mm ($\square$), and $d$ = 0.3 mm ($\bigtriangledown$). Marker shapes shown in parentheses are used in all figures in the following to represent the corresponding grain size.


\subsection{Cohesion Controlled Granular Material (CCGM)}
\label{subsec:sec2_2_CCGM}

\begin{table}
\begin{center}
\begin{tabular}{||m{6em}||m{1.5cm}||m{1cm}||m{1cm}||m{2cm}|m{2cm}|m{2cm}|m{2cm}||} 
\hline\hline
$d$ (mm) & Source & $b$ (nm) & $W_{\%}$ & $\langle\phi\rangle$ & $\mu$ & $\tau_{\rm y}$ (Pa) & \textbf{$\mathrm{Co}$} \\ [0.5ex]
\hline\hline
$0.34 \pm 0.04$ & - & 0 & 0  & $0.60 \pm 0.01$ & $0.35 \pm 0.02$ & 0 & 0 \\
$1.08\pm\,0.10$ & - & 0 & 0 & $0.55 \pm 0.01$ & $0.35\pm 0.03$ & 0 & 0\\ 
\hline
$1.08\pm\,0.10$ & CCGM & 116  & - & $0.60 \pm 0.01$ & $0.33 \pm 0.05$ & $35 \pm 20$ & $2.2 \pm 1.3$\\ 
$0.70\pm\,0.05$ & CCGM & 106  & - & $0.56\pm 0.01$ & $0.43 \pm 0.04$ & $28 \pm 10$ & $3.0 \pm 1.0$\\ 
\hline
$1.08\pm\,0.10$ & Wet & -  & 0.5$\%$ & $0.55 \pm 0.02$ & $0.35 \pm 0.05$ & $111 \pm 13$ & $7.5 \pm 0.9$\\ 
$0.70\pm\,0.05$ & CCGM & 310 & - & $0.57\pm 0.01$ & $0.27 \pm 0.07$ & $79 \pm 31$ & $8.1 \pm 3.2$\\ 
\hline
$0.49 \pm 0.04$ & Wet & - & 0.5$\%$ & $0.51\pm 0.04$ &  $0.46\pm 0.16$ & $133 \pm 42$ & $20 \pm 6.1$\\
\hline \hline
\end{tabular}
\caption{Physical details of the cohesive grains used in this study. 
The static properties ($\langle\phi\rangle$, $\mu$, $\tau_{\rm y}$) are determined using the inclined shear cell. {Solid fraction $\phi = V_{\rm solid}/ V_{\rm tot}$ is measured for every trial and then averaged for a given cohesive granular material. The coefficient of friction, $\mu$, and the yield stress $\tau_{\rm y}$ are through the results of the shear cell using the Mohr-Coulomb failure criterion, as shown in Fig. \ref{fig:Fig2_Characterization}.}}
\label{table:table1_Shearcell}
\end{center}
\end{table}

We first {create} inter-particle cohesion {through} a polymer coating {on} the grains, {following the approach of Gans \textit{et al.}} \cite{2020_gans}. The cohesion is induced by a polyborosiloxane (PBS) coating on the glass {beads}, made of polydimethylsiloxane (PDMS) cross-linked with Boric Acid. More details on the manufacturing can be found in Ref. \cite{2020_gans}. Grains made using this method are dry, and separating and re-attaching them does not affect their stickiness. We prepare three batches of CCGM, two corresponding to {relatively} small cohesions (made by coating $b \approx 100$ nm on grains {of diameter} $d=1.1$ mm or $d=0.7$ mm), and a third CCGM with moderate cohesion {for which} grains are made of a thicker coating ($b \approx 300$ nm and $d=0.7$ mm). {The physical properties of} these grains {can be found} in Table \ref{table:table1_Shearcell}. The magnitude of the cohesive force is controlled by tuning the average coating thickness, $b$. 
\smallskip

{The precise physical origin of the cohesion observed for the CCGM is still not fully understood \cite{2020_gans}. Indeed, for a short contact duration, Gans \textit{et al.} described the cohesive force using a capillary model, from which one can infer that cohesion arises from the establishment of bridges between the grains. In this case, if the coating thickness $b$ is below a certain value (which depends on the roughness of the grain), increasing it leads to an increase of the number of contact points, hence the overall cohesion force. All our considered CCGM for this study are in this regime.} {If $b$ exceeds a threshold, the roughness of the grains is screened by the coating, and additional complications may also be noticed when such grains are kept in contact for a large time, which seems to be due to an aging phenomenon such as the entanglement of PDMS polymer chains \cite{2020_gans}.} To ensure {that aging does not play a significant role in our experiments}, the same collapse experiments were tested by waiting a few seconds after preparing the initial column and another by waiting an hour. Both experiments led quantitatively to the same results. 
\smallskip

While Gans \textit{et al.} \cite{2020_gans} provided an \textit{ad-hoc} relationship between the cohesive force $F_{\rm c}$ and the coating thickness $b$, this relationship was not straightforwardly applicable to {describe} our grains with a Bond number. {It is indeed possible to define a} granular Bond number $\mathrm{Bo}$ as the ratio between the {weight of} a particle {and} the cohesive force {using the \textit{ad-hoc} expression of Ref. \cite{2020_gans}}. {Nevertheless, }in our experience, two manufactured CCGM with supposedly similar Bond numbers produced {very} different {levels of} cohesion for the range of grain sizes considered in this study. Thus, additional tests are required to determine the actual forces or the yield stresses to characterize CCGM. Section \ref{sec:sec3_Cohesion} discusses our tests to determine the yield stresses.


\subsection{Wet Grains in the Pendular State}
\label{subsec:sec2_3_Wet}

The second source of cohesion {between the grains is induced by} capillary bridges of deionized (DI) water \cite{herminghaus2005dynamics,2006_mitarai} {with surface tension $\gamma_{\ell} = 72$ mN.m$^{-1}$ at ambient temperature}. When the liquid bridges are small compared to the diameter of the grain $d$, the capillary force between two spherical particles is proportional to $\gamma_{\ell} \, d$ \cite{2006_mitarai}. We define the mass liquid content in an unsaturated {wet} granular material as $W_{\%} = m_{\mathrm{water}} / m_{\mathrm{particles}}$ \cite{2013_artoni,2021_li,2022_li}. If all the liquid content in such a granular medium is confined to liquid bridges, the material is said to be in the pendular state \cite{2004_kohonen, 2006_mitarai}. This {situation} also corresponds to the state of mixture between phases where the material has the largest cohesion \cite{2006_richefeu}, and maximum shear and elastic moduli \cite{2007_moller}.
\smallskip

To produce these grains, we first remove any moisture content by drying the grains in an oven. Then water is weighed and added {to} the grains {once they} have cooled {altogether in a closed container}, following Li \textit{et al.} \cite{2021_li}. The closed container containing grains and water is placed on a mixer for 5 minutes to ensure the water is homogeneously mixed with the grains before moving into the {column}. In preliminary experiments, the grains are moved between containers to estimate the water content lost in moving grains into the experimental apparatus by adhesion to the {surface of the container} (approx. 2$\%$ by mass). Consequently, this much extra water mass is added to prepare our cohesive grains \cite{2021_li, 2006_richefeu}.
\smallskip

Artoni \textit{et al.} \cite{2013_artoni} showed that the effects of such cohesion on the final morphology of the 2D channelized deposit can be rationalized by {a group of dimensionless numbers,} $\mathrm{Bo}^{-1} \, {W_{\%}}^{2/3}$ consisting of the granular Bond number, $\mathrm{Bo}$ and the liquid content $W_{\%}$. The granular Bond number is defined as \cite{2013_artoni}: \begin{equation}
\label{eq:Bond_cap}
\mathrm{Bo} = \frac{F_{\rm w}}{F_{\rm c}} = \frac{\rho \, g \, d^2} {\gamma_{\ell}} \, , 
\end{equation} where $\rho$ is the particle density and {and $g$ is the acceleration due to gravity}. Li \textit{et al.} \cite{2022_li} also rationalized the final morphology of unchannelized collapse of rectangular columns using this dimensionless group. For our study, we seek to minimize {any viscous dissipation, and consequently, the attempt is made to keep $W_{\%}$ as small as possible while ensuring a homogeneous mixture of grains and fluid.} Altogether, for grains in the pendular state, we report experiments {with two grain sizes ($d=0.5$ mm or $1.1$ mm) and for which $W_{\%}=0.5\%$}. Increasing grain size corresponds to an increase in the Bond number {and}, thus, {to} a less cohesive system.
\smallskip

\section{Characterization of Cohesive Strength}
\label{sec:sec3_Cohesion}

\begin{figure}
\centering
\includegraphics[width=0.95\textwidth]{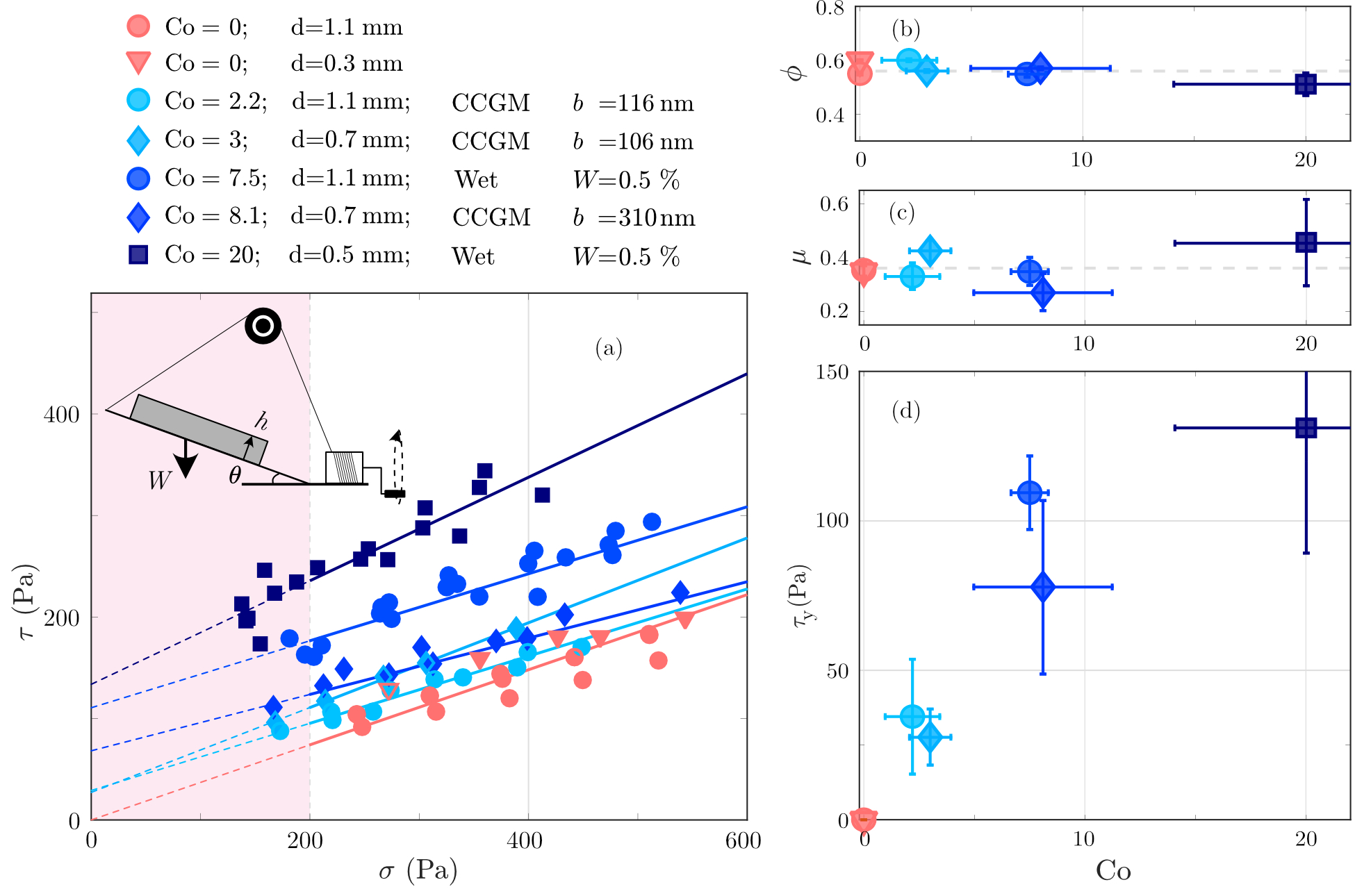}
\caption{(a) Inset: Schematic of the inclined plane setup. By changing the height $h$, the weight $W$ can be changed and controlled. Main figure: {Shear and normal stresses $\tau$ and $\sigma$, respectively,} corresponding to each failure experiment for all the tested grains. The best fits of the yield curves, {which are shown in solid lines}, are used to estimate the yield stress $\tau_{\rm y}$ and the friction coefficient $\mu$ using the Mohr-Coulomb failure criterion {from Eq. \eqref{eq:Coulomb_Cohesion}}. (b) Measured volume fraction $\phi$ from the shear cell experiments for our range of grains {as a function of $\mathrm{Co}$ [as defined in Eq. \eqref{eq:Co_Stress}]}. $\phi$ is averaged from all the trials, and the standard deviation {from} this average is the associated uncertainty. The evolution of the friction coefficient $\mu$ and the yield stress $\tau_{\rm y}$ as a function of $\mathrm{Co}$ are shown in (c) and (d), respectively. The uncertainties in both subfigures show the variation of fits.}
\label{fig:Fig2_Characterization}
\end{figure}

As we shall see later, our experiments suggest that the {final deposits} of some {cohesive columns made with} grains {using} different sources of cohesion {and} grain sizes {are} very {close}. For instance, glass beads of $d$=1.1 mm and $W_{\%} = $ 0.5$\%$ {exhibit very similar behavior to that of a} CCGM of diameter $d$=0.7 mm and $b\approx 300$ nm. Independent measurements of cohesive yield stresses are carried out using a shear cell to compare the macroscopic behavior of different materials. A similar approach was considered by Santomaso \textit{et al.} \cite{2018_santomaso} to compare collapses of rectangular columns made of wet spherical beads and calcium carbonate (coarse sand), where no straightforward expression could be {derived for} the cohesion force to be used in the Bond number description.
\smallskip

The shear cell is used to evaluate the cohesion for each granular system {based on} the Mohr-Coulomb failure criterion {(the simpler model for granular plasticity)} \cite{1992_nedderman, 2013_andreotti}. The yield stress described through this criterion is a measure of internal failure within a granular bulk. At a point of failure, the {state} of stresses on the {failing} plane must overcome the effects of friction, having accounted for cohesion. Measurements of shear and normal stress configurations at failure are used to estimate the yield locus. Within this framework, a cohesive material will yield when the applied shear $\tau$ reaches: \begin{equation}
\label{eq:Coulomb_Cohesion}
\tau = \mu\, \sigma + \tau_{\rm y}\, , 
\end{equation} where $\sigma$ is the normal stress, $\mu$ is the coefficient of internal friction, and $\tau_{\rm y}$ is the cohesive yield strength of the material. Richefeu \textit{et al.} \cite{2006_richefeu} use a rectangular shear cell to characterize the shear strength of wet granular materials. Here, similar to Gans \textit{et al.} \cite{2020_gans}, we use an inclined plane apparatus.
\smallskip

We start with a cohesive granular bed {placed in a cell of} fixed basal surface $S$ (length 20 cm, width 10 cm), whose height $h$ is varied between tests. The bottom surface of the cell is coated with a layer of grains of the same diameter as the grains being tested {\cite{2020_gans}}. Since the failure occurs on this {bottom} plane, this coating ensures {the grains do not slide}. The cell is initially filled with the cohesive material. The top {plane} remains open, and {coincides with the cell's height}. The bed is weighed before an experiment, and since the internal volume and weight of the empty cell are known, we obtain the volume fraction $\phi$ {of the tested material}. The filled cell is placed on an inclined plane{, as illustrated by the} schematic of the setup shown in the inset of Fig. \ref{fig:Fig2_Characterization}(a). The downstream side of the box is opened such that the grains are {free to fall without any restriction} as the system is inclined at a continuous rate. As the angle $\theta$ increases, the magnitude of normal stresses on the bottom surface of the cell decreases, {while} the shear stresses increase until the material yields at a critical angle, $\theta_{\mathrm c}$. Since we measure the initial weight of the bulk of grains, $W = \phi\, \rho \, g\,h\, S$, at the point of failure, the normal stress on the bottom surface is $\sigma = W \, {\rm cos}\, \theta_c / S$ and the shear stress is $\tau = W \, {\rm sin}\, \theta_c / S$. By changing the height $h$ of the initial bed of grains, we change $W$ and thus the angle at which the avalanche is triggered. {While local collapses do occur at the opening of the downstream side, the plane used is large
enough that the area affected is small compared with bulk. For each set of grains}, at least five different heights are used. The results are reported in Fig. \ref{fig:Fig2_Characterization}(a), {with} the best fits {of the data using Eq. \eqref{eq:Coulomb_Cohesion}} also shown. The intercept is $\tau_{\rm y}$, and the slope is a measure of the coefficient of internal friction, $\mu$. The yield locus displays some non-linearity when the normal stresses are small, typically for $\sigma < 200$ Pa here. A similar behavior was reported by Richefeu \textit{et al.} \cite{2006_richefeu}, and following their approach, the fits do not include these points. {Additionally, it should be noted that this shear cell technique is ill-suited to measure failure for cohesionless grains, as the free surface always approximately exhibits the angle of repose. This explains the wide range of data for our cohesionless points, and for simplicity, $\tau_{\rm y} = 0$ is enforced for this case.}
\smallskip

The values of $\tau_{\rm y}$ measured from {these} tests are used to classify our cohesive granular materials on a common scale. Santomaso \textit{et al.} \cite{2018_santomaso} suggested that $\tau_{\rm y}$ can be used to explain the effects of cohesion on collapse. However, {anticipating the discussion given in the next section,} the cases reported in Figs. \ref{fig:Fig5_TopView}(b) and \ref{fig:Fig5_TopView}(c) correspond to two different $\tau_{\rm y}$, despite producing quantitatively similar final deposits for all aspect ratios. The yield loci for these two grains are marked in indigo on Fig. \ref{fig:Fig2_Characterization}(a) for the wet grains ($\bigcirc$) and the CCGM ($\Diamond$). {It shows that} the yield stress is not adequate independently to describe the effects of cohesion on collapse. The additional effects of grain size on the final morphology{, which is demonstrated later in the present study,} were likely unnoticed by Santomaso \textit{et al.} \cite{2018_santomaso} since coarse sand and glass beads of roughly equal size were used in their experiments.
\smallskip

Rather than considering the effects of cohesion at the scale of an individual bond, $\tau_{\rm y}$ measures the cohesion within a static bulk of cohesive grains. Gans \textit{et al.} \cite{2020_gans} have shown for CCGM that the cohesive force scales as $F_c \sim \tau_{\rm y}\,d ^{2}$, where $d$ is the grain diameter. In the case of wet grains, \citet{1997_pierrat} derived a relation from Rumpf's model for the tensile strength $\sigma_{\rm t}$ in terms of the attractive force {$F_c$} between grains and their diameter and find $F_{c} \sim \sigma_{\rm t}\, d ^{2}$, which is further elucidated by Richefeu \textit{et al.} \cite{2006_richefeu} for additional particle properties. $\sigma_{\rm t}$ is a measure of the tensile strength and represents the resistance of individual bonds to traction \cite{2018_santomaso}. The tensile strength scales as the cohesive yield stress, with a numerical prefactor related to the internal angle of friction \cite{1992_nedderman}. Most importantly, this implies that {one} can use $\tau_{\rm y}$ to estimate the average cohesion force at the scale of the grain as $\langle F_{c} \rangle \sim \tau_{\rm y}\, d^2$, independent of the source of cohesion for these two cases, and measured only through its bulk failure. {Consequently}, we define a cohesion number using this average description of forces: \begin{equation}
\label{eq:Co_Stress}
\mathrm{Co} \equiv \frac{\langle F_{\rm c} \rangle}{\langle F_{\rm w}\rangle} \cong \frac{\tau_{\rm y}\, d^2}{\phi\,\rho\, g\, d^3} = \frac{\tau_{\rm y}}{\phi\,\rho\, g\, d}\,, 
\end{equation} Here, we reversed the numerator and denominator {compared to the definition of the Bond number}, such that a more cohesive granular bulk corresponds to a larger {value of} $\mathrm{Co}$. Consequently, {we expect} $\mathrm{Co} \sim \mathrm{Bo}^{-1}$. {Since we are considering the average cohesion force in the bulk, we similarly consider the bulk weight at the scale of the grain, $\langle F_{\rm w}\rangle \sim \rho_{b} \, g \, d^3$, where $\rho_b = \phi \, \rho$ is the bulk density.} Despite not knowing the numerical prefactors to resolve the forces individually, the expression here is readily comparable for all our cohesive grains since $\tau_{\rm y}$ and $\phi$ {can be} measured experimentally. This description can also be interpreted as a balance of relevant length scales, as {it is discussed later} in section \ref{sec:sec5_Discussion}.
\smallskip

{The values of $\mathrm{Co}$ for all tested grains are compiled in Table \ref{table:table1_Shearcell}}. Altogether, the cohesive effects can be broadly categorized as the following: (i) cohesionless grains; (ii) small cohesion - CCGM; (iii) moderate cohesion - our most cohesive CCGM, and the less cohesive wet grains; (iv) large cohesion - our most cohesive wet grains. Since the bulk cohesion description {also} accounts for the change in grain size, we refer only to {the cohesive number} $\mathrm{Co}$ for brevity in what follows. Fig. \ref{fig:Fig2_Characterization}(b) shows the {average} volume fraction $\langle\phi\rangle$ measured {per grain species from measurements} prior to the start of each test in the shear cell. Error bars on this plot are the standard deviations of the measured $\phi$ for each kind of cohesion. For our range of $\mathrm{Co}$ and grain size, {observed $\phi$ for all the grains are similar}, {with an overall mean value} $\langle\phi\rangle = 0.56 \pm 0.02$, {corresponding to loose packings}. A slightly lower $\langle\phi\rangle$ is observed{, however,} for the largest cohesion considered here. In Fig. \ref{fig:Fig2_Characterization}(c) and \ref{fig:Fig2_Characterization}(d), the values of $\mu$ and $\tau_{\rm y}$ are shown, respectively, {as a function of} $\mathrm{Co}$. {Error bars, in these plots, show uncertainties from the fit.} The friction coefficient $\mu$ is roughly constant ($\langle\mu\rangle = 0.36 \pm 0.01$) for our range of experiments{, however a wider scatter of the data is noticed for larger cohesion. In Fig. \ref{fig:Fig2_Characterization}(d),} the two grains with moderate {cohesions ($\mathrm{Co}$ = 7.5 and 8.1)} correspond to different yield stresses, but have a similar overall $\mathrm{Co}$. {In addition,} the two wet grains {(for which $\mathrm{Co}$ = 7.5 and 20)} correspond to similar yield stresses but are quite different qualitatively. This difference is, again, captured in {the definition of the cohesive number $\mathrm{Co}$} since smaller $d$ results in a larger $\mathrm{Co}$.
\smallskip

\section{Results}
\label{sec:sec4_Results}


\subsection{Dynamics: {Phenomenology} of 3D Collapse}
\label{subsec:sec4_1_Dynamics}

\begin{figure}
\centering
\includegraphics[width = \textwidth]{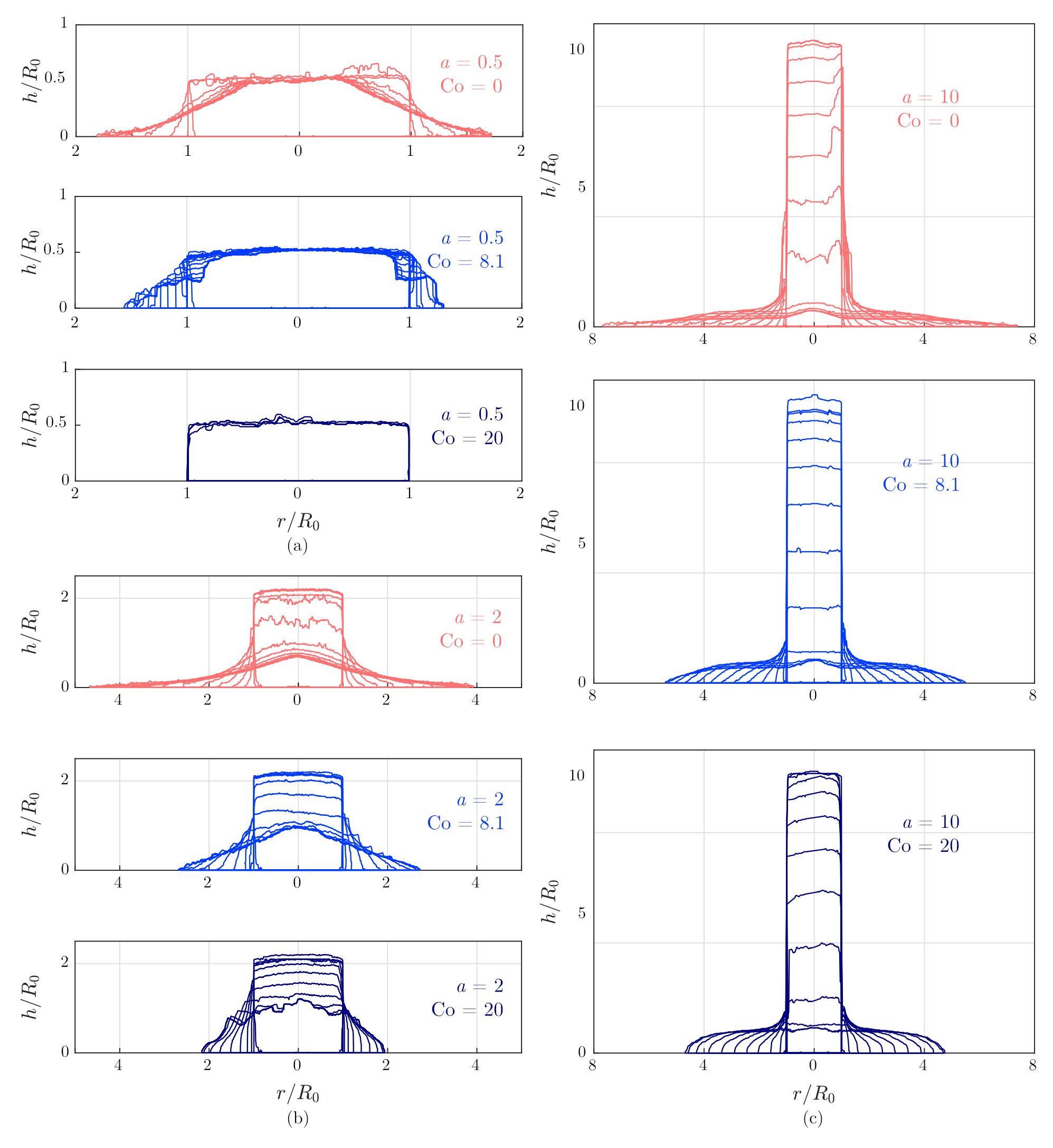}
\caption{Side profiles of the cylindrical column collapsing axisymmetrically for various initial aspect ratios, $a$, and cohesions, $\mathrm{Co}$, taken every 0.03 s. (a) $a = 0.5$, (b) $a = 2$ and (c) $a = 10$. For each $a$, we report collapses with cohesionless grains, \textit{i.e.}, $\mathrm{Co}$ = 0, a medium cohesion with $\mathrm{Co}$ = 8.1 and with our most cohesive grains with $\mathrm{Co}$ = 20. The corresponding videos are available in supplemental materials.}
\label{fig:Fig3_TimeSeries}
\end{figure}

\begin{figure}
\centering
\includegraphics[width = \textwidth]{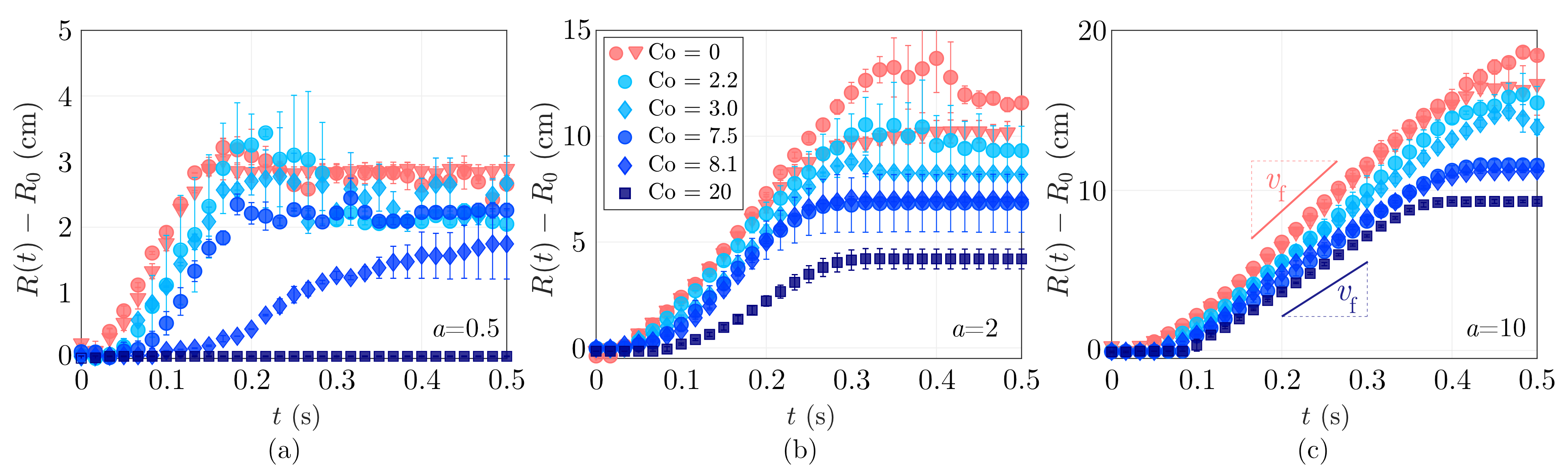}
\caption{Time evolution of the spreading radius for the axisymmetric collapse of cylindrical columns after triggering the withdrawal of the confining cylinder. (a) $a = 0.5$, (b) $a=2.0$, and (c) $a=10$ for our full range of cohesion.
These cases correspond to the evolutions shown in Fig. \ref{fig:Fig3_TimeSeries} with additional intermediate cohesion values.}
\label{fig:Fig4_Transient}
\end{figure}

{In what follows, we investigate the collapse of cohesive granular columns, by varying both the initial aspect ratio $a = H_0 / R_0$ and the value of the cohesive number $\mathrm{Co}$.} For all aspect ratios considered in this study, once released, {unstable} columns spread unconfined for {approximately} 0.5 s before coming to rest. To illustrate the collapsing column, a selection of the side profiles is shown in Fig. \ref{fig:Fig3_TimeSeries}. We show three cases, for which $H_0 < R_0$ ($a=0.5$), $H_0 \gtrsim R_0$ ($a=2.0$) and $H_0 \gg R_0$ ($a=10.0$) in Figs. \ref{fig:Fig3_TimeSeries}(a)-(c), respectively. {For each $a$, a} cohesionless collapse (\textit{i.e.}, $\mathrm{Co}=0.0$), a moderate cohesion ($\mathrm{Co}=8.1$), and our most cohesive grains tested ($\mathrm{Co}=20$) {are shown}. Frames captured every successive 1/30 s after the cylindrical shell begins lifting are presented. Complete videos of these collapses are available in supplemental materials. In the cohesionless case, the removal of the cylinder causes the outermost layer of grains to be disturbed by friction on the cylinder wall, which {is highlighted by} roughness of {significant} amplitude in some early profiles. Still, as these grains collapse, the deposits exhibit a smooth final surface. {However,} roughness is seen on the surface of the final deposits of cohesive grains at the end of the dynamics. {Major} roughness is also {present} for small $a$ in the cohesive cases, where agglomerates fall off from the sides of the column. The collapsing column spreads less for more cohesive grains, as expected. 
\smallskip

From the side profiles, we extract the time evolution of the granular front {at the base of the column, $R(t) - R_0$}. In Figs. \ref{fig:Fig4_Transient}(a)-(c), we show the radial evolution corresponding to $a =$ 0.5, 2.0, and 10, respectively, for all {cohesion} tested. Error bars indicate the difference between the left and right profiles seen by the camera, the average of which is taken to be the radial front. In these plots, $t=0$ s is denoted as the time corresponding to the first frame where the cylindrical enclosure begins to be lifted. Time on these axes thus shows the time since the grains are free to start moving. For Fig. \ref{fig:Fig4_Transient}(a), where $a=0.5$, the most cohesive case $\mathrm{Co}=20$ shows no radial change at all, as the column is stable [as illustrated, for instance, in Fig. \ref{fig:Fig3_TimeSeries}(a)]. For $\mathrm{Co}=8.1$, the column initially holds steady and then collapses after some delay. Particularly for the larger aspect ratios, as shown in Figs. \ref{fig:Fig4_Transient}(b) and \ref{fig:Fig4_Transient}(c), the evolution of the radial front is well described by a straight line (\textit{i.e.} a constant velocity), with brief acceleration and deceleration {phases}. The acceleration and deceleration {stages} at the start and the end of the spreading are even less pronounced for more cohesive grains. The slope is used to determine the {average} velocity of the front, $v_f$, as marked in Fig. \ref{fig:Fig4_Transient}(c), which appears smaller for collapses with larger cohesion. 

\subsection{Morphology of Final Deposits}
\label{subsec:sec4_2_Morphology}

\begin{figure}
\centering
\includegraphics[width = 1.0\textwidth]{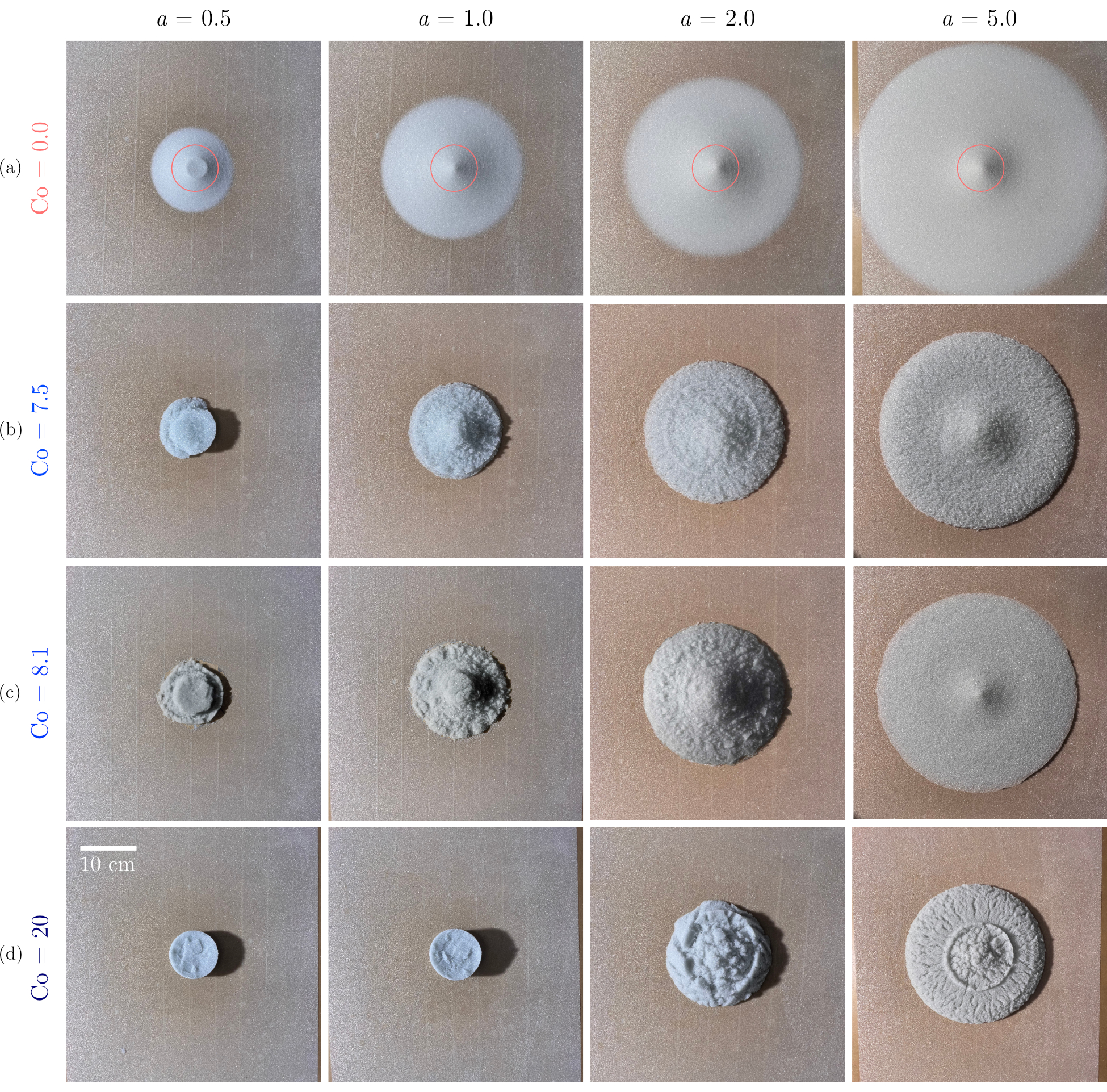}
\caption{A top view of the relaxed piles for a selection of aspect ratios $a$ and cohesive numbers $\mathrm{Co}$. (a) $\mathrm{Co}=0$, with a trace of the initial cylindrical column of radius $R_0 =$ 4.2 cm shown. (b) and (c) correspond to {moderate cohesions for} wet grains ($\mathrm{Co}$ = 7.5) and CCGM ($\mathrm{Co}$ = 8.1), respectively, and lead to similar results. (d) $\mathrm{Co}$ = 20 corresponds to the most cohesive grains used in this study. A common scale bar is shown, corresponding to 10 cm.}
\label{fig:Fig5_TopView}
\end{figure}

\begin{figure}
\centering
\subfigure[]{\includegraphics[width = 0.49\textwidth]{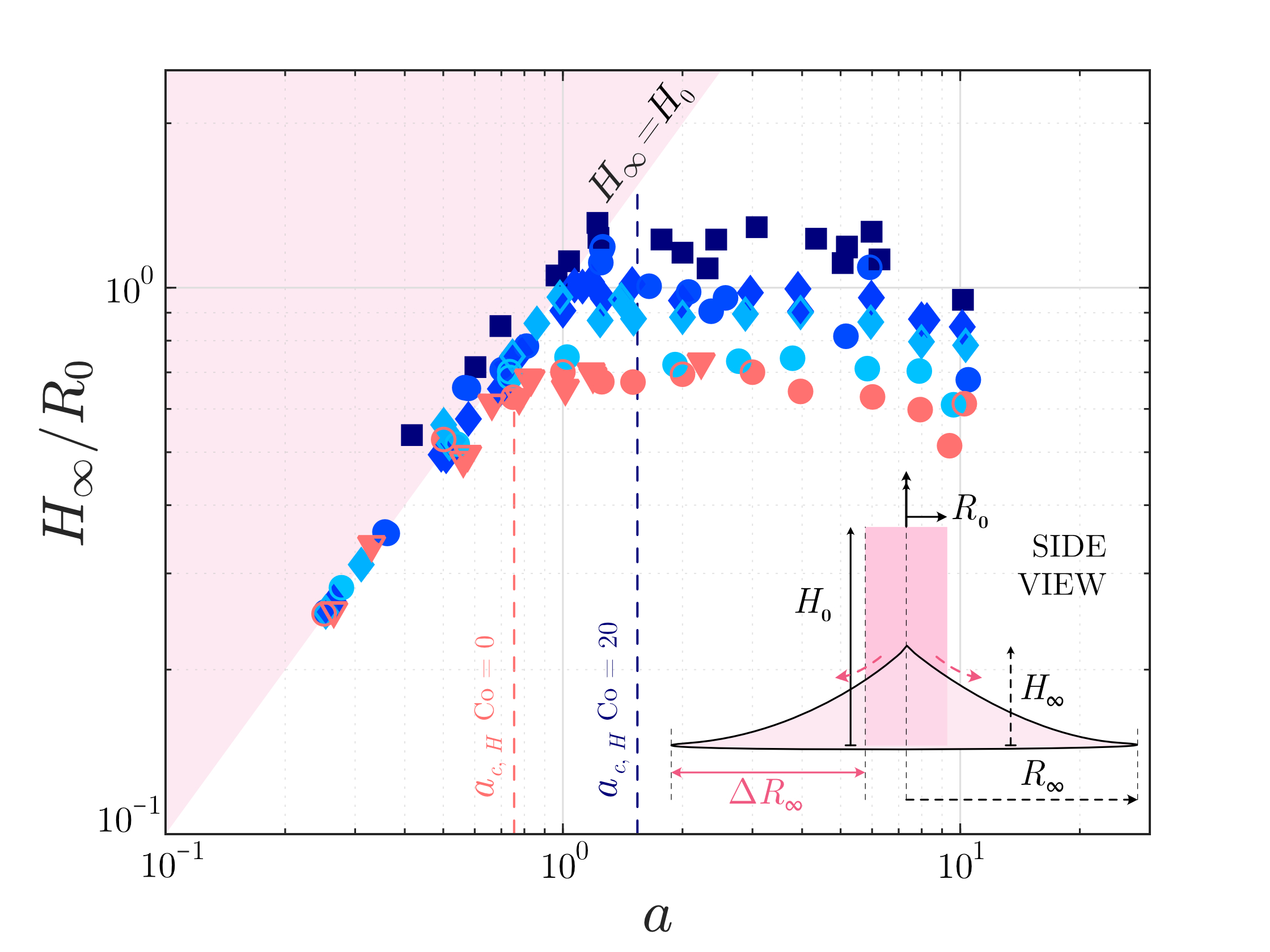}}
\subfigure[]{\includegraphics[width = 0.49\textwidth]{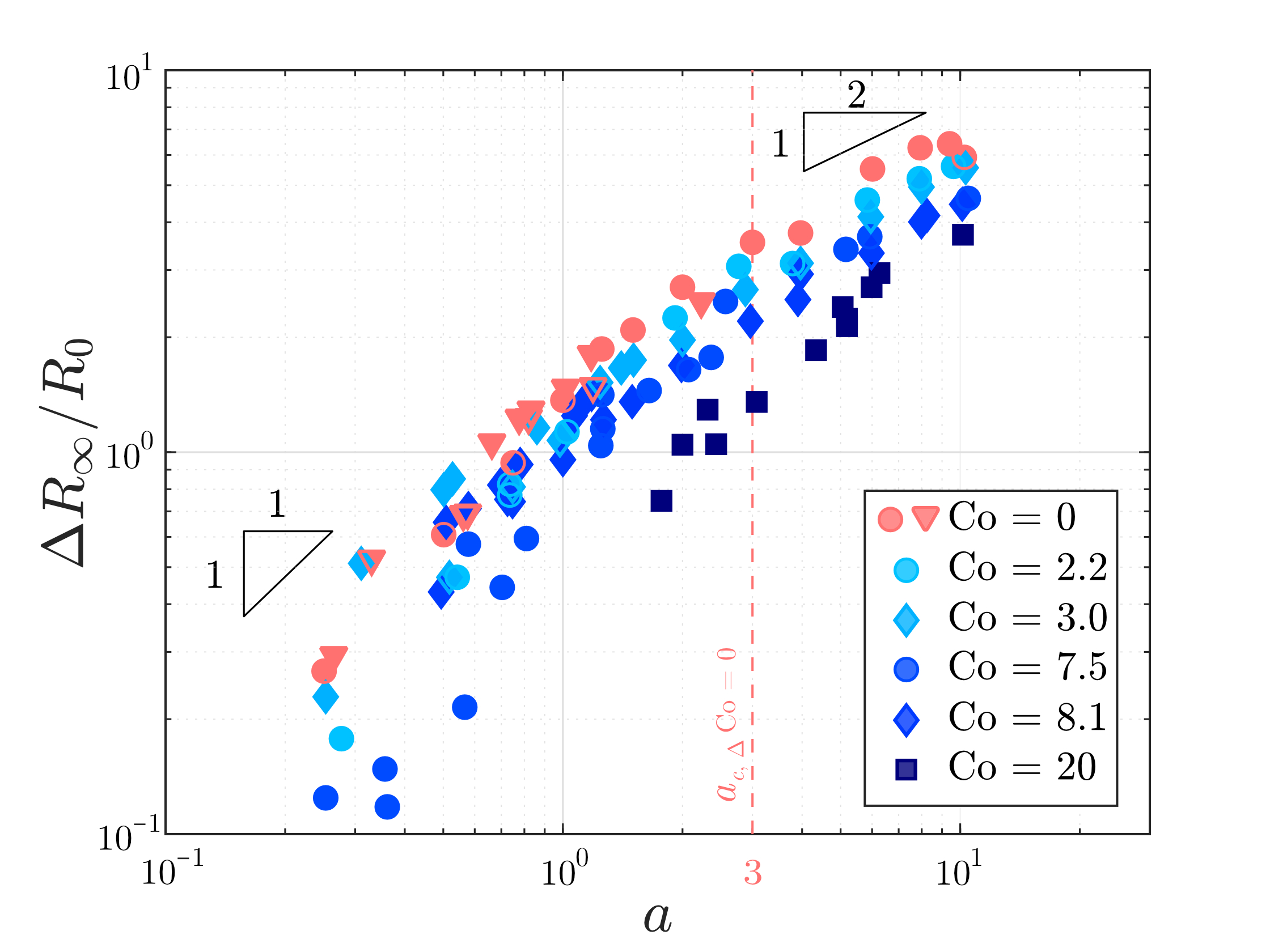}}
\caption{Final extent of the deposit for a 3D cylindrical column of grains collapsing axisymmetrically for a range of initial aspect ratios $a$ and cohesions $\mathrm{Co}$. (a) Final rescaled height $H_{\infty} / R_0$ as a function of $a$ for all {investigated values of} $\mathrm{Co}$. The critical aspect ratio ${a_{c, \, H}}$ separates the two trends as formulated in Eq. \eqref{eq:1}. ${a_{c, \, H}}$ for cohesionless grains and {for} the most cohesive grains tested are both marked by dashed vertical lines. The shaded region corresponds to $H_{\infty} > H_0$, which is not physically possible since the column cannot stretch. (b) Evolution of the rescaled radial runout, $\Delta R_{\infty}/R_0 = R_{\infty}/R_0 - 1$ as a function of $a$ for the tested $\mathrm{Co}$. ${a_{c, \, \Delta}} = 3$ is shown by the dashed line separating the two power-law trends for the cohesionless grains. Stable columns, \textit{i.e.}, columns for which $R_{\infty} = R_0$ do not appear on this plot.}
\label{fig:Fig6_FinalMorphology_3D}
\end{figure}

\begin{figure}
\centering
\subfigure[]{\includegraphics[width = 0.49\textwidth]{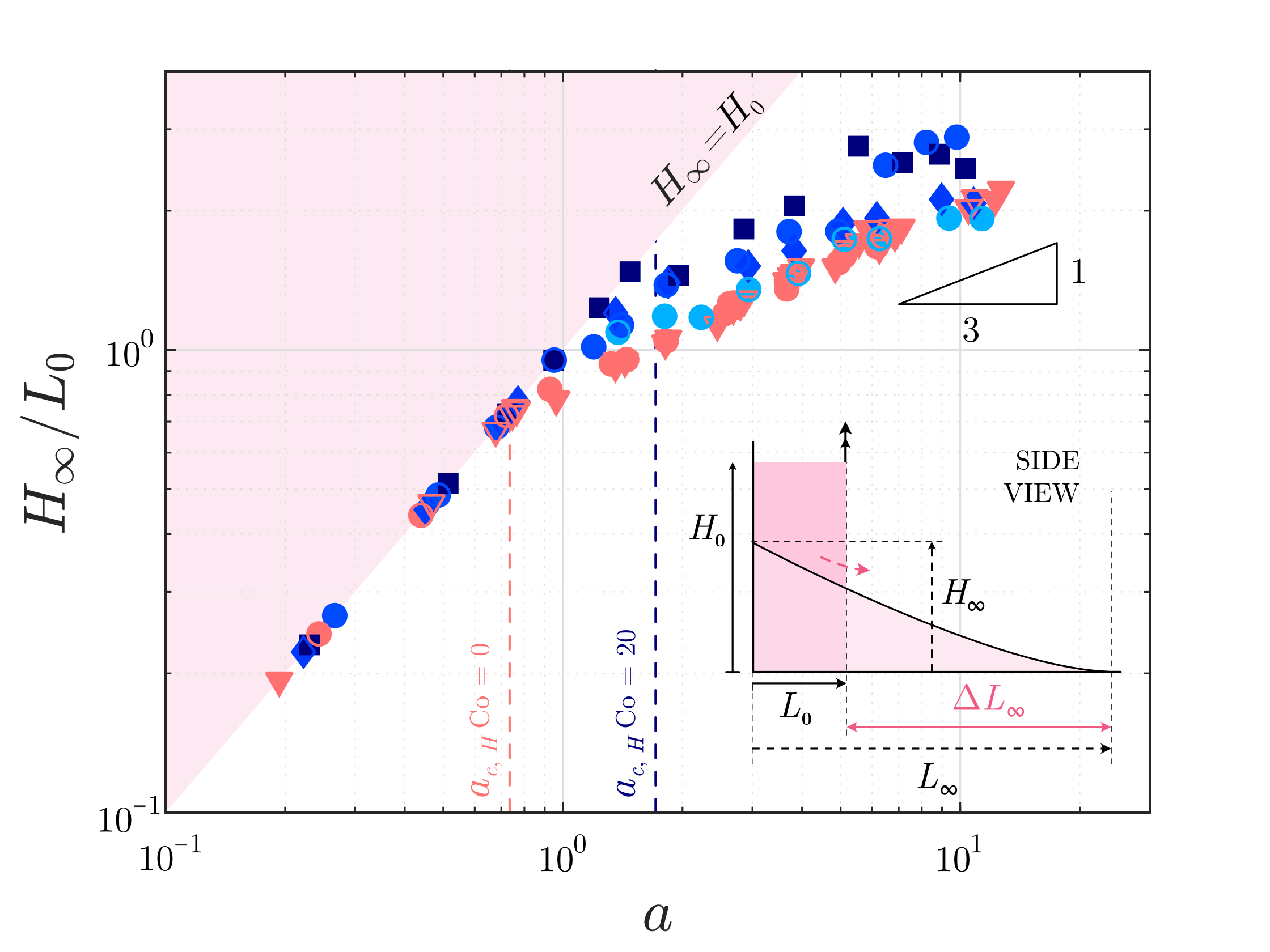}}
\subfigure[]{\includegraphics[width = 0.49\textwidth]{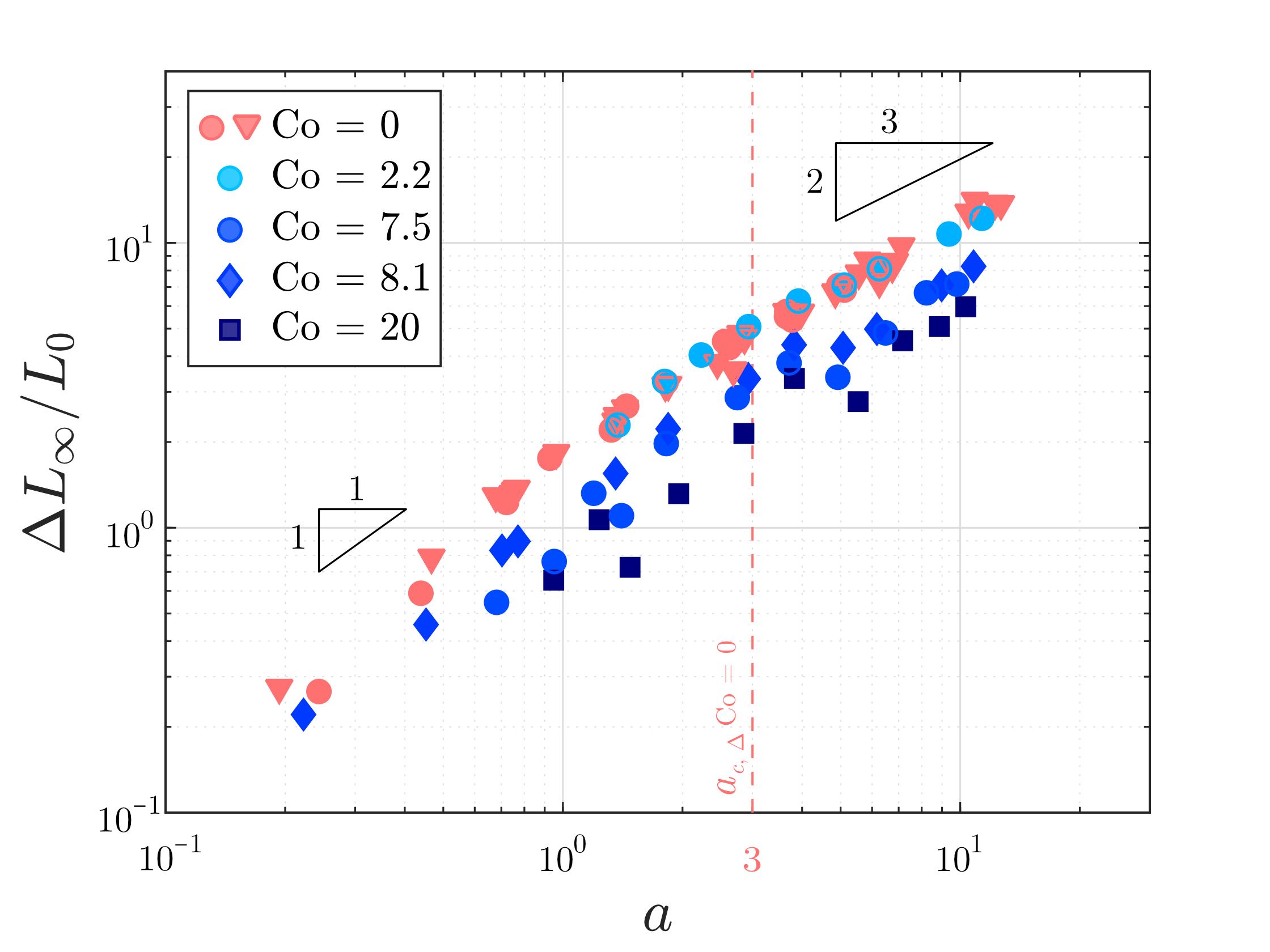}}
\caption{Final extent of the deposit for a 2D rectangular column of grains collapsing in a constrained channel for a range of initial aspect ratios $a$ and cohesions $\mathrm{Co}$. (a) Final rescaled height $H_{\infty} / L_0$ as a function of $a$ for all $\mathrm{Co}$ tested here. The critical aspect ratio ${a_{c, \, H}}$ separates the two trends as formulated in Eq. \eqref{eq:2}. ${a_{c, \, H}}$ for cohesionless grains and the most cohesive grains tested are both marked by dashed vertical lines. The shaded region corresponds to $H_{\infty} > H_0$, which is not physically possible since the column cannot stretch. (b) Evolution of the rescaled axial runout, $\Delta L_{\infty}/L_0 = (L_{\infty} - L_0)/L_0$ as a function of $a$ for the tested $\mathrm{Co}$. ${a_{c, \, \Delta}} = 3$ is shown by the dashed line separating the two power-law trends for the cohesionless grains. Stable columns, \textit{i.e.}, columns for which $L_{\infty} = L_0$ do not appear on this plot.}
\label{fig:Fig7_FinalMorphology_2D}
\end{figure}

Examples of the axisymmetric deposit, once the granular collapse has finished, are shown in Fig. \ref{fig:Fig5_TopView} for a few cohesion numbers, $\mathrm{Co}$, and initial aspect ratios, $a$. {For} each image, the initial radius of the column {is} $R_0 = 4.2$ cm, and all {pictures} share the same scale. Fig. \ref{fig:Fig5_TopView}(a) shows collapses of cohesionless grains, and the circumference of the confining cylinder is penciled in to show its relative size. Figs. \ref{fig:Fig5_TopView}(b)-(d) show collapses  of $\mathrm{Co}$ = 7.5, 8.1 and $\mathrm{Co}$ = 20 grains, respectively. For aspect ratio $a = 1$, {the final height} $H_{\infty} = H_0$ for these cohesions, while $H_{\infty} < H_0$ for $\mathrm{Co}$ = 0, suggesting cohesion affects the {critical aspect ratio separating} partially and fully collapsing regimes, ${a_{c, \, H}}$.
\smallskip

In the first regime, \textit{i.e.}, for $a < {a_{c, \, H}}$, the presence of cohesion can lead to two different kinds of collapse. In some combination of parameters, particularly large $\mathrm{Co}$ and small $a$, collapse do not occur at all (\textit{e.g.}, $a$ = 0.5 and 1 for $\mathrm{Co}$ =20), akin to stable columns in sandcastles. When cohesive grains do collapse in this regime, the collapse is irregular and often not symmetric - agglomerates of different sizes fall off the column in blocks. These observations are reminiscent of recent experiments for the unchannelized collapse of rectangular columns of wet grains in the pendular state {performed} by Li \textit{et al.} \cite{2022_li}. For large aspect ratios, where $a > {a_{c, \, H}}$, the final deposits display some memory of the initial cylinder on the top surface as a crown-like shape on the final deposit (e.g., Co = 7.5 and 8.1 and $a$ = 2.0). For large cohesion and aspect ratios (e.g., Co = 20 and $a$ = 5.0), the final deposit displays clear crack-like striations, which appear {during spreading. The significant roughness, and sometimes cracks, visible on the morphology for the cohesive cases seems to be the consequence of the collapse dynamics, as it is more pronounced for higher aspect ratios (\textit{i.e.}, for high inertia of the grains during collapse), and may depend on the source of cohesion. Indeed, a closer inspection of Fig \ref{fig:Fig5_TopView}(b) and \ref{fig:Fig5_TopView}(c) reveals that the spatial distribution of roughness differs between wet grains and CCGM, so does the corona-like shape on the final deposits for $a=2$: While it is almost continuous for wet grains ($\mathrm{Co}=7.5$), in the case of CCGM ($\mathrm{Co}=8.1$) its consists of a ring of singular agglomerates. The investigation of these aspects is beyond the scope of the present study as it would require finer measurements of the surface roughness.}
\smallskip

In all cases, cohesion reduces the overall spread of a deposit compared to cohesionless columns of the same initial dimensions. The final deposits also display some roughness on the free surface as a consequence of these materials being able to {sustain} tensile stresses for some grain layers below the free surface \cite{1992_nedderman, 2021_abramian}. The effects of cohesion on the surface roughness on the final deposit of a 2D collapse were recently studied using numerical simulations by Abramian \textit{et al.} \cite{2021_abramian}, where more cohesive grains typically lead to a rougher final deposit.
\smallskip

The final height $H_{\infty}$ and the runout $\Delta R_{\infty} = R_{\infty} - R_0$ (for the axisymmetric case) or $\Delta L_{\infty} = L_{\infty} - L_0$ (for the rectangular geometry) of the deposits are measured from the profiles once the grains have stopped spreading. The pile reaches its relaxed state approximately 0.5 s after the cylinder begins to be lifted and does not undergo any noticeable changes in the next few minutes. Lube \textit{et al.} \cite{2004_lube}, and Lajeunesse \textit{et al.} \cite{2004_lajeunesse} have shown these length scales of the final deposit to depend on the initial aspect ratio $a = H_0/R_0$ {(or $H_0/L_0$)} of the column for both these geometries {in the case of} cohesionless grains. The influence of $a$ on $H_\infty$ and $\Delta R_\infty$, when {those two lengths are} rescaled by the initial radius $R_0$ (or width $L_0$ in two dimensions), is shown to be captured by empirical piece-wise power laws of the form \cite{2005_lajeunesse}:
{\allowdisplaybreaks
\begin{alignat}{4}
  \label{eq:1}
	\frac{H_\infty}{R_0} = & \left\{ \begin{array}{ll}
	\displaystyle a\\[8pt]
	\displaystyle \alpha
	\end{array}\right.
    & \begin{array}{ll}
	\displaystyle \ \mathrm{for\ }  a \lesssim {a_{c, \, H}},\\[8pt]
	\displaystyle \ \mathrm{for\ }  a \gtrsim {a_{c, \, H}},
	\end{array}
    \qquad
	\frac{\Delta R_\infty}{R_0} = & \left\{
	\begin{array}{ll}
	\displaystyle \beta a\\[8pt]
	\displaystyle \gamma a^{1/2}
	\end{array}\right.
    & \begin{array}{ll}
	\displaystyle \ \mathrm{for\ }  a \lesssim {a_{c, \, \Delta}},\\[8pt]
	\displaystyle \ \mathrm{for\ }  a \gtrsim {a_{c, \, \Delta}},
	\end{array}
\end{alignat}}
\noindent for the axisymmetric setup, and
{\allowdisplaybreaks
\begin{alignat}{4}
  \label{eq:2}
	\frac{H_\infty}{L_0} = & \left\{ \begin{array}{ll}
	\displaystyle a\\[8pt]
	\displaystyle \delta a^{1/3}
	\end{array}\right.
    &\begin{array}{ll}
	\displaystyle \ \mathrm{for\ }  a \lesssim {a_{c, \, H}},\\[8pt]
	\displaystyle \ \mathrm{for\ }  a \gtrsim {a_{c, \, H}},
	\end{array}
    \qquad
	\frac{\Delta L_\infty}{L_0} = & \left\{
	\begin{array}{ll}
	\displaystyle \varepsilon a\\[8pt]
	\displaystyle \zeta a^{2/3}
	\end{array}\right.
    &\begin{array}{ll}
	\displaystyle \ \mathrm{for\ }  a \lesssim {a_{c, \, \Delta}},\\[8pt]
	\displaystyle \ \mathrm{for\ }  a \gtrsim {a_{c, \, \Delta}},
	\end{array}
\end{alignat}
}\noindent for the rectangular case. Note that the critical aspect ratios ${a_{c, \, H}}$ and ${a_{c, \, \Delta}}$ differ between the two scalings \eqref{eq:1} and \eqref{eq:2} {according to Lajeunesse \textit{et al.} \cite{2005_lajeunesse}}. The values for the numerical prefactors $\alpha$, $\beta$, $\gamma$, $\delta$, $\varepsilon$, and $\zeta$, as well as those of the critical aspect ratios ${a_{c, \, H}}$ and ${a_{c, \, \Delta}}$ marking the transitions between the collapse regimes, are {known to depend} on the material and frictional properties of the grains \cite{2005_balmforth,2021a_man}. Figs. \ref{fig:Fig6_FinalMorphology_3D} and \ref{fig:Fig7_FinalMorphology_2D} show the rescaled {length scales $H_\infty/R_0$ and $\Delta R_\infty/R_0$ (resp. $H_\infty/L_0$ and $\Delta L_\infty/L_0$)} of the final deposits of all our experiments for the axisymmetric and rectangular geometries, respectively. Special attention is brought to the trends of $\mathrm{Co}$ = 7.5 and 8.1, which correspond to two granular systems having different sources of cohesion and grain sizes yet lead to similar results for the entire range of tested $a$. This is {further} evidence for {the relevance of using an} average cohesion description since this suggests {a bulk description such as that the cohesive number} $\mathrm{Co}$ can capture the macroscopic effects of cohesion on collapse.
\smallskip

In Figs. \ref{fig:Fig6_FinalMorphology_3D}(a) and \ref{fig:Fig6_FinalMorphology_3D}(b), we show measurements of the rescaled height and runout of the final deposit of the axisymmetric collapse as a function of $a$ for the range of cohesive grains considered. In the case of the ratio $H_{\infty}/R_0$  presented in Fig. \ref{fig:Fig6_FinalMorphology_3D}(a), two regimes are {clearly visible and are} separated by a critical aspect ratio ${a_{c, \, H}}$. For $a < {a_{c, \, H}}$, some central portion of the column does not participate in the collapse, and consequently, $H_{\infty} = H_0$. We determine ${a_{c, \, H}}$ as the average of the largest $a$ for which $H_{\infty} = H_0$ and the smallest $a$ for which $H_{\infty} < H_0$. Another method to estimate ${a_{c, \, H}}$ would be to consider the intersect of a plateau with the line {$H_\infty/R_0=a$}. However, this approach could lead to some discrepancies if the resulting values do not follow a plateau. For cohesionless grains (\textit{i.e.} $\mathrm{Co}=0$), experiments were done with grains of size $d$ = 1.1 mm ($\bigcirc$) and $d$ = 0.3 mm ($\bigtriangledown$) and ${a_{c, \, H}}$ is found to be 0.71, in close agreement with Lajeunesse \textit{et al.} \cite{2004_lajeunesse} who found ${a_{c, \, H}}$ = 0.74 with similar sized grains for the axisymmetric collapse. The transition between the two regimes, ${a_{c, \, H}}$, {then} increases for more cohesive grains, e.g., ${a_{c, \, H}}$ for $\mathrm{Co}=20$ is found to be 1.50, and is marked in Fig. \ref{fig:Fig6_FinalMorphology_3D}(a) with a dashed line. The variation of ${a_{c, \, H}}$ for our range of {cohesive number} $\mathrm{Co}$ is shown in Fig. \ref{fig:Fig8_Prefactors_3D_2D}(a). Since dry columns cannot stretch, $H_{\infty} = H_0$ demarcates a physical limit on the final height: the {corresponding} non-physical region $H_{\infty} > H_0$ is shaded in the figure. For our largest cohesion ($\mathrm{Co}=20$), minor stretching is noted as cohesion allows the column to sustain extensional stresses when the cylinder is removed. For $a > {a_{c, \, H}}$, all the relaxed piles exhibit a plateau value of $H_{\infty}/R_0$, independent of the initial aspect ratio and larger for more cohesive grains. The value of this plateau is denoted as $\alpha$ in the following and obtained by fitting $H_{\infty}/R_0$ for all the values of $a > {a_{c, \, H}}$ for each cohesive material. For $\mathrm{Co}=0$, $\alpha$ is found to be $0.66 \pm 0.03$ from the fit, in {reasonable} agreement with Lajeunesse \textit{et al.} \cite{2004_lajeunesse} who found $\alpha$ = 0.74. A full list of ${a_{c,\, H}}$ and $\alpha$ for all the tested cohesive grains is provided in Table \ref{table:table2_LajScal}.
\smallskip

Similarly, $\Delta R_{\infty}/R_0$ displays two trends with the aspect ratio $a$, as shown in Fig. \ref{fig:Fig6_FinalMorphology_3D}(b). The separation between the two {regimes} defines a second critical aspect ratio ${a_{c, \, \Delta}}$. Following Lajeunesse \textit{et al.} \cite{2004_lajeunesse}, we use ${a_{c, \, \Delta}} =3$ for our cases as well since the behavior is systematic beyond this threshold for all our grains {and because the effects of cohesion on $a_{c, \, \Delta}$ are less obvious than for the relative final height}. For $a \lesssim {a_{c, \, \Delta}}$, a linear trend is observed such that $\Delta R_{\infty}/ R_0 = \beta a$, where $\beta$ is a numerical prefactor, as has been suggested by Lube \textit{et al.} \cite{2004_lube} {based on scaling arguments}. For $a \gtrsim {a_{c, \, \Delta}}$, $\Delta R_{\infty} /R_0 = \gamma a^{1/2}$, where $\gamma$ is another numerical prefactor. Fitting our cohesionless data to these power laws leads to comparable values of  $\beta$ and $\gamma$ to those found in previous studies \cite{2004_lube,2005_lajeunesse,2014_warnett}. The presence of cohesion reduces the radial spread of the pile in both regimes, analogous to observations reported in other geometries \cite{2013_artoni, 2021_li,2022_li,2023_gans}. A {scattering} of the data is observed for small aspect ratios $a < {a_{c, \, \Delta}}$. Indeed, in these cases, we observe agglomerates of grains falling off the column, resulting in some associated randomness with such collapses. {However, it is observed that, at first order, cohesion only shifts the curves, with no major modification of the followed trends in the investigated range of $\mathrm{Co}$, except for the most cohesive case where $\mathrm{Co}=20$. This suggests that, overall, cohesion essentially affects the numerical prefactors of the scalings given by Eqs. \eqref{eq:1}. In what follows, we thereby assume that this is effectively the case, so that even for cohesive grains we will apply similar fits as those given by Eq. \eqref{eq:1}.} The prefactor values of $\beta$ and $\gamma$ {obtained by doing so} are provided in Table \ref{table:table2_LajScal} for all our grains. {This} assumption, {and its relevance,} are discussed in more detail in Appendix \ref{sec:AppA_Fits}.
\smallskip


In Figs. \ref{fig:Fig7_FinalMorphology_2D}(a) and \ref{fig:Fig7_FinalMorphology_2D}(b), the respective measurements of $H_{\infty}/L_0$ and $\Delta L_{\infty}/L_0$ from the channelized rectangular collapses are shown. For the rescaled final height in Fig. \ref{fig:Fig7_FinalMorphology_2D}(a), the two pieces are again clearly separated by a critical aspect ratio, ${a_{c, \, H}}$. ${a_{c, \, H}}$ for $\mathrm{Co}$=0 is found to be 0.84, in close agreement with previous experiments in this geometry {\cite{2005_lajeunesse}}. Similar to the axisymmetric case, the value of ${a_{c, \, H}}$ is larger for more cohesive grains, as shown in {Fig. \ref{fig:Fig8_Prefactors_3D_2D}(e)}. For $a > {a_{c, \, H}}$, the final re-scaled height {was fitted by a power law of the form} $H_{\infty}/R_0$= $\delta\,a^{1/3}$, where $\delta$ is a numerical prefactor \cite{2005_lajeunesse}. {The value of $\delta$ for $\mathrm{Co} = 0$ is found from fitting to be $\delta_0 = 0.91 \pm 0.02$, and increases with $\mathrm{Co}$}. Likewise, for this geometry, we suggest that the effects of cohesion are {mainly} captured by changes in the prefactors for the scaling laws of the rectangular collapse, {a point that is also discussed} in Appendix \ref{sec:AppA_Fits}.
\smallskip

The rescaled channelized {runout distances} are shown in Fig. \ref{fig:Fig7_FinalMorphology_2D}(b). Similar to the axisymmetric case, {$a_{c, \, \Delta}$ is kept fixed at a value of ${a_{c, \, \Delta}}$} = 3, following Lajeunesse \textit{et al.} \cite{2005_lajeunesse}. Fitting the piece-wise power law from \eqref{eq:2} for cohesionless grains, we obtain $\varepsilon_0$ and $\zeta_0$ to be 1.71 $\pm$ 0.01 and 2.43 $\pm$ 0.07 respectively, in good agreement with literature on this geometry. {In summary, cohesion} reduces the {spreading of all} columns. Table \ref{table:table4_LajScal_2D} summarizes the details of the fits for the range of cohesive grains tested.

\section{Discussion}
\label{sec:sec5_Discussion}


\subsection{Effects of Cohesion on Final Morphology}
\label{subsec:sec5_1_DataCollapse}

Altogether, inter-particle cohesion affects the velocity of the spreading grains, the overall spread, the shape of the final deposit, and the roughness of the free surface. Effects of cohesion on the final deposit across both collapse geometries are found {to depend on the cohesive} number, $\mathrm{Co}$, which relates the macroscopic cohesion to grain-scale properties. In this section, the effects of cohesion on the final morphology are {quantified and related to the value of} $\mathrm{Co}$.
\smallskip

\begin{figure}
\centering
\includegraphics[width=\textwidth]{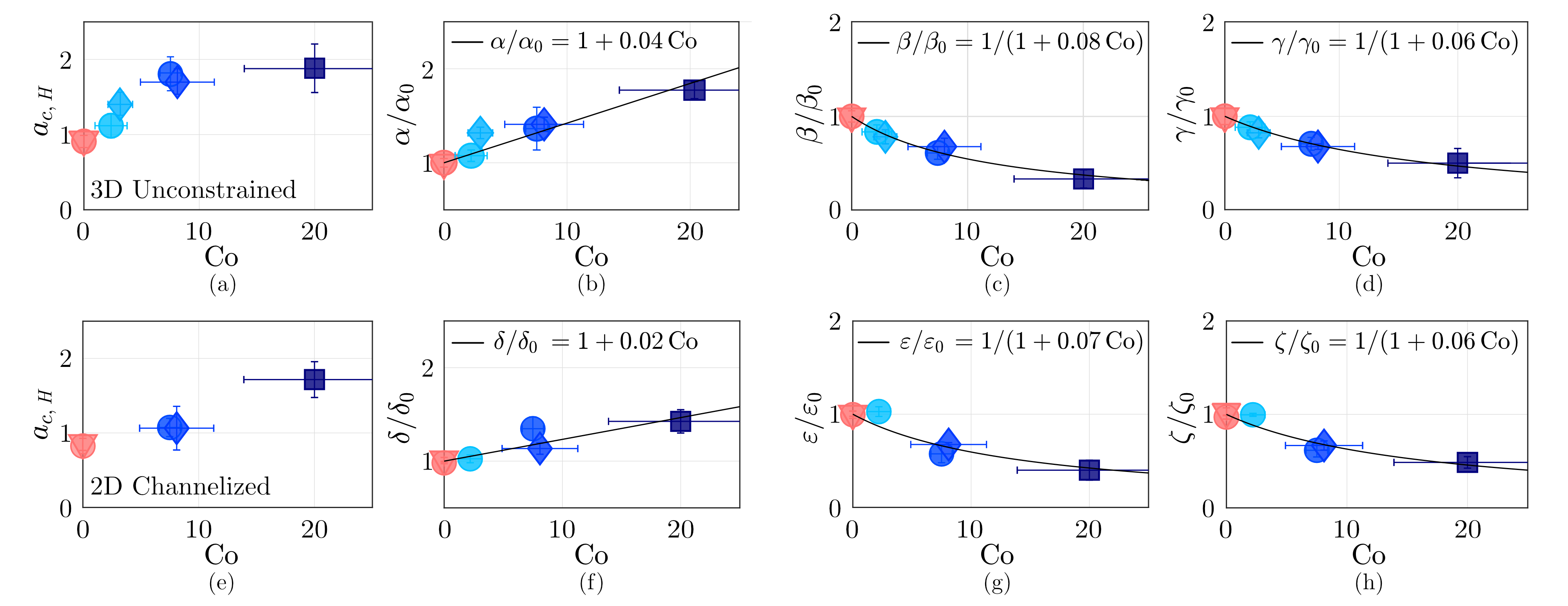}
\caption{(a) and (e) Critical aspect ratio ${a_{c, \, H}}$ separating the two trends for the {rescaled final} height of the relaxed pile, $H_{\infty} /R_0$, for each cohesion in the 3D and 2D geometry, respectively. (b) - (d) The effects of cohesion on the numerical prefactors $\alpha,\, \beta$ and $\gamma$ as defined in Eq. \eqref{eq:1} for the 3D collapses, normalized by their respective values for $\mathrm{Co}$=0. (f) - (h) The effects of cohesion on the numerical prefactors $\delta,\, \varepsilon$ and $\zeta$ as defined in Eq. \eqref{eq:2} for the 2D channelized collapses, also normalized by their respective values for $\mathrm{Co}$=0. Simple fits to capture the relative effects of cohesion on the final deposit are shown within each plot.}
\label{fig:Fig8_Prefactors_3D_2D}
\end{figure}

The power laws {of Eqs. \eqref{eq:1} and \eqref{eq:2}, derived by Lajeunesse \textit{et al.} \cite{2005_lajeunesse},} are used to describe the final deposit. The corresponding {numerical} coefficients associated with each cohesion are listed in Table \ref{table:table2_LajScal} for the axisymmetric geometry and {in} Table \ref{table:table4_LajScal_2D} for the rectangular case. Figs. \ref{fig:Fig8_Prefactors_3D_2D}(a) and \ref{fig:Fig8_Prefactors_3D_2D}(e) show that for both 3D and 2D geometries, larger cohesion leads to larger critical aspect ratios ${a_{c, \, H}}$. In Fig. \ref{fig:Fig8_Prefactors_3D_2D}(b), values of $\alpha$ for our cohesive grains are normalized by $\alpha_0$, where $\alpha_0 = \alpha_{\mathrm{Co}=0}${, and shown as a function of $\mathrm{Co}$}. A {good collapse of the data is observed, which motivates the use of a} linear fit applied to the {experimental} points to capture the evolution of the plateau $\alpha$ with $\mathrm{Co}$. Similarly, for the 2D case {presented} in Fig. \ref{fig:Fig8_Prefactors_3D_2D}(f), the evolution of $\delta$ normalized by $\delta_0 = \delta_{\mathrm{Co}=0}$ {as a function} of $\mathrm{Co}$ is shown along with {its corresponding best} linear fit. Altogether, despite the difference in the power laws, the effects of cohesion on the rescaled final height for {$a > a_{c, \, H}$} are found to be similar across both these geometries:
\allowdisplaybreaks
\begin{alignat}{4}
  \label{eq:alpha_delta}
	\alpha = \alpha_0 (1 + 0.04\,\mathrm{Co}) \,;
    \qquad
	\delta = \delta_0 (1 + 0.02\,\mathrm{Co}) \,.
\end{alignat} The effects are slightly more pronounced in the case of the axisymmetric collapse but {overall they remain} of similar magnitude. This is likely due to the difference in $H_{\infty}$ between the geometries: In 3D, the peak is supported only by grains below, while in 2D, the presence of a confining boundary wall causes differences.
\smallskip

In the case of the runout {distance}, similar steps are used to explicitly determine the effects of cohesion on numerical prefactors $\beta$ and $\gamma$ for the axisymmetric case, and $\varepsilon$ and $\zeta$ for the rectangular case, {corresponding to} the {different collapse} regimes. Since cohesion reduces the spread, these {numerical} prefactors reduce with increased cohesion. The critical aspect ratio delineating the pieces of the power laws is taken as {$a_{c, \, \Delta} = 3$} for both geometries and all considered cohesions. For small aspect ratios, i.e., {in the linear regimes where} {$a < a_{c, \, \Delta}$}, the effects of cohesion on the runout are shown in Figs. \ref{fig:Fig8_Prefactors_3D_2D}(c) and \ref{fig:Fig8_Prefactors_3D_2D}(g) for the axisymmetric and rectangular geometries, respectively. In both cases, prefactors $\beta$ and $\varepsilon$ from the fit are normalized by {their cohesionless values} $\beta_0$ and $\varepsilon_0$ and shown for the range of {investigated} $\mathrm{Co}$. A fit is shown in a solid line for both plots showing similar effects of cohesion {as for the final} height, with the caveat that cohesion reduces the overall spread, {so that the fits now take the form} \begin{alignat}{4}
  \label{eq:beta_epsilon}
	\beta = \frac{\beta_0}{1 + 0.08\,\mathrm{Co}} \,;
    \qquad  
	\varepsilon = \frac{\varepsilon_0}{1 + 0.07\,\mathrm{Co}} \,.
\end{alignat} Using scaling arguments, Lube \textit{et al.} \cite{2005_lube} have shown that the re-scaled spread is linearly related to the initial aspect ratio $a$ for {$a < a_{c, \, \Delta}$}, for both 2D and 3D geometries. Our results {suggest} that the effects of cohesion are also observed to be similar across both geometries in this regime{, as $\beta/\beta_0 \approx \varepsilon/\varepsilon_0$}. Furthermore, even for {$a > a_{c, \, \Delta}$}, the effects of cohesion on the spreading are comparable across the geometries. {Indeed,} while the difference in geometry results in different powers for the dependence on $a$, the effects of cohesion are found to {mainly} affect the numerical prefactors in a similar way. Figs. \ref{fig:Fig8_Prefactors_3D_2D}(d) and \ref{fig:Fig8_Prefactors_3D_2D}(h) show the change in the fitted prefactors for the 3D and 2D geometry, respectively, for the range of cohesion {investigated}. Similar to the small aspect ratio cases, two fits are {also reported in Figs. \ref{fig:Fig8_Prefactors_3D_2D}(d) and \ref{fig:Fig8_Prefactors_3D_2D}(h)}. These are found to be: 
\begin{alignat}{4}
  \label{eq:gamma_zeta}
	\gamma = \frac{\gamma_0}{1 + 0.06\,\mathrm{Co}} \,;
    \qquad  
	\zeta = \frac{\zeta_0}{1 + 0.06\,\mathrm{Co}} \,.
\end{alignat} for the axisymmetric and rectangular cases, respectively. Two conclusions can be drawn from these: First, this confirms that the effects of cohesion on the final spread are comparable in 2 or 3 dimensions for the range of cohesions tested {in this study}. This also shows that the influence of cohesion does not vary much between the two pieces of the piece-wise scaling for the rescaled runout.
\smallskip

\begin{figure}
\centering
\includegraphics[width = \textwidth]{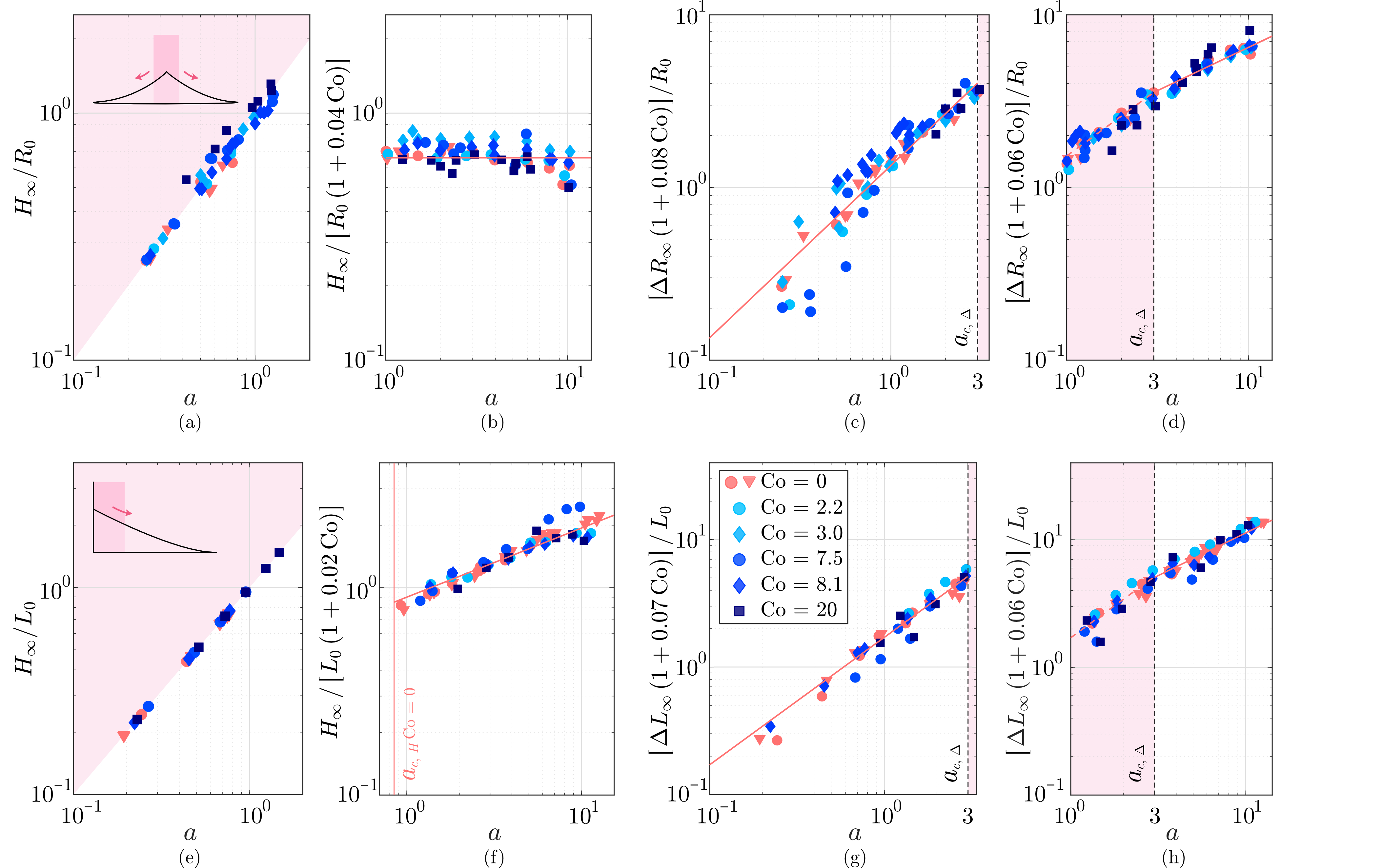}
\caption{Data collapse for the final deposit for the full range of initial aspect ratios $a$ and cohesions $\mathrm{Co}$. (a)-(d) {Rescaled lengths of the final deposit {for the 3D configuration}, renormalized using the fits} shown in Figs. \ref{fig:Fig8_Prefactors_3D_2D}(b)-(d){, as a function of the aspect ratio}. Similarly, (e)-(h) {presents the rescaled lengths of the final deposit for the 2D channelized collapse, renormalized using the fits shown in Figs. \ref{fig:Fig8_Prefactors_3D_2D}(f)-(h)}. In each piece of the power law, the solid line marks the best fit for cohesionless collapses, onto which the rest of the data is {thereby} collapsed.}
\label{fig:Fig9_DataCollapse_3D_2D}
\end{figure} 

At first glance, the contribution of $\mathrm{Co}$ may {seem} small since the numerical coefficients {of the obtained scalings} are small, but this is purely an artifact of how $\mathrm{Co}$ is defined. For instance, a column of $\mathrm{Co} \approx 15$ would result in half the total spread of a cohesionless column of the same initial dimensions. $\mathrm{Co} \approx 30$ would reduce the total runout by a factor of 3, etc. Using such fits, we now have {a quantitative estimate of the influence} of cohesion for the final height and runout, for the two geometries {considered} and {the investigated range of cohesion}. {Our data is replotted in Figs. \ref{fig:Fig9_DataCollapse_3D_2D}(a)-(d) and \ref{fig:Fig9_DataCollapse_3D_2D}(e)-(h), by removing the contribution of cohesion using the fits from Fig. \ref{fig:Fig8_Prefactors_3D_2D}}. {It is observed that} the data collapses onto the cohesionless case reasonably well, with {cohesionless} fits marked explicitly in solid lines. Altogether, the primary effects of the tested cohesions on the final spread of collapsed columns across geometries are captured by {Eqs. \eqref{eq:alpha_delta}, \eqref{eq:beta_epsilon}, and \eqref{eq:gamma_zeta}. It should be emphasized that these relations are independent of the source of cohesion, and can be rationalized instead by bulk measurements of yield stress, $\tau_{\rm y}$, bulk solid fraction, $\phi$, average grain size, $d$, and density, $\rho$, as discussed in Sec. \ref{sec:sec3_Cohesion}.}


\subsection{An Alternative Interpretation of the Cohesive Number}
\label{subsec:sec5_2_AltCo}

The bulk cohesion number $\mathrm{Co}$ has been defined in Eq. \eqref{eq:Co_Stress} at a macroscopic scale using independent yield stress measurements $\tau_{\rm y}$ {for the considered} cohesive granular material. $\tau_{\rm y}$ is used to estimate the average cohesive force at the scale of the grain as $\langle F_c \rangle \sim \tau_{\rm y}\, d^2$. However, inter-particle cohesion is often also described in terms of {a characteristic length scale}, $\ell_c$ \cite{1992_nedderman,2020_gans,2020_abramian,2020_onoDitBiot, 2003b_courrech_du_pont}. From scaling arguments,  $\ell_c \sim \tau \, (\rho \,g)^{-1}$, where $\tau$ is some characteristic stress scale. In this description of cohesive granular materials, this length is analogous to the capillary length scale of fluids \cite{2020_onoDitBiot}. Following our earlier definitions, we {choose here to} include $\phi$ in {the definition of} $\ell_c$, such that $\ell_c \equiv \tau_{\rm y} \, (\phi\,\rho\,g)^{-1}$. {This decision is made in order to consider the bulk density $\rho_b = \phi \rho$}. The bulk cohesion number from Eq. \eqref{eq:Co_Stress} can then be re-written as:\begin{equation}
\label{eq:Co_final}
\mathrm{Co} \equiv \frac{\tau_{\rm y}}{\phi\,\rho \, g\, d} = \frac{ \ell_c}{d}\,, 
\end{equation} {Through this approach, the cohesive number} can be interpreted as the ratio {between} the characteristic length scale introduced by cohesion {and the typical grain size}. This last description {was also recently} proposed by Abramian \textit{et al.} \cite{2020_abramian} as a macroscopic description {accounting for the influence of interparticle cohesion during the collapse of rectangular columns of cohesive grains}. Further, their framework uses the force between particles for cohesive contact forces rather than only capillary forces. This nuance is important here since the cohesions considered in this study are not induced solely through capillary forces. Using the Rumpf-Richefeu model \cite{2006_richefeu}, the macroscopic description of cohesion can be related to the Bond number, as shown by \cite{2020_abramian}:\begin{equation}
\label{eq:microMarco_RR}
\frac{\ell_c}{d} = \frac{\mu\, \phi Z}{4}\, \mathrm{Bo}^{-1}  \, 
\end{equation} where $\mu$ is the coefficient of static friction, $Z$ is the average number of contacts, and the Bond number $\mathrm{Bo}\equiv F_w/F_c$ is the ratio of forces at the scale of an individual particle. In our case, since $\phi$ is included within the definition of $\mathrm{Co}$ in Eq. \eqref{eq:Co_Stress}, we get $\mathrm{Co} = (\mu \, Z / 4)\,  \mathrm{Bo}^{-1}$. Since we do not make explicit measurements of $Z$ or the force {required} to break an individual cohesive bond $F_c$, this expression {can} not {be} verified here. However, with $\mathrm{Co} \sim \mathrm{Bo}^{-1}$, our results can be qualitatively compared to the experiments of collapses {using} wet grains \cite{2013_artoni, 2018_santomaso, 2021_li} or CCGM \cite{2023_gans}, where the $\mathrm{Bo}$ is known. In the results from these previous studies for rectangular columns, cohesion leads to larger final heights of the deposit and a lower axial runout{, while these quantities are} found to be systematically related to $\mathrm{Bo}^{-1}$. {This enforces the idea that} the effects of cohesion described in {this} study are {indeed} systematically related to a macroscopic definition of {the cohesive number} $\mathrm{Co}${,  which has been shown here to be valid for two different geometries of cohesive granular collapse.}

\section{Conclusion}
\label{sec:sec6_Conclusion}
In this study, the collapse of cohesive granular columns under the effect of gravity has been considered for two geometries: unconfined {(3D)} and channelized {(2D)}. Two different sources of cohesion have been investigated: Due to capillary bridges, or induced by a polymer coating. In general, the presence of larger cohesion results in a slower spread, a smaller runout, an overall rougher final free surface{, and a larger} final height of the {obtained} deposits. For the range of cohesion considered here, the {typical dimensionless lengths associated with the final morphologies} scale with the initial aspect ratio with similar power laws {as} those established for cohesionless collapses. The effects of cohesion on the final deposits are then rationalized using the yield strength of a particular cohesive system, altogether comparing an average cohesion force to the weight of {a} grain. Since the measurements of the yield strength are done considering bulk failure, this method of characterizing cohesion can be used to compare the two sources of cohesion, independent of the {details regarding forces} at the scale of the individual bonds. This characterization of cohesion is used to explain the effects of small cohesion on collapse{, which at first order impacts only} the numerical prefactors to the established power laws. Finally, these experiments provide validation for {the approach followed by} Abramian \textit{et al.} \cite{2020_abramian} and the Rumpf-Richefeu model \cite{2006_richefeu}, relating bulk effects of cohesion to particle-scale effects.
\smallskip

{The influence} of cohesion on the final deposit is explained in the form of first-order modifications to the cohesionless {scalings}. For a full picture, a larger range of cohesions {would need to be investigated}, {which could be achieved by} connecting {numerical} simulations such as those performed by Langlois \textit{et al.} \cite{2015_langlois} to a macroscopic description such as the one provided here. While dimensional analysis requires the inclusion of the density of the particles $\rho$ in the macroscopic description, this parameter {was not varied} in {the present} study and {thus} requires {further experimental attention}. The two sources of cohesion considered here are capable of reforming after being broken, {however how the present approach} would have to be modified if cohesion was brittle remains to be {investigated}. Lastly, whether such a macroscopic description can be readily applied to systems where yield strength results from entanglement and geometry of particles also remains to be explored.

\clearpage
\appendix

\section{Additional Details on Fits}
\label{sec:AppA_Fits}

\begin{table}
\begin{center}
\begin{tabular}{||m{6em}|m{6em}||m{2cm}|m{2cm}||m{2cm}|m{2cm}||} 
\hline\hline
\textbf{$\mathrm{Co}$} & $d$ (mm) & ${a_{c,\, H}}$ & $\alpha$ & $\beta$ & $\gamma$ \\ [0.5ex]
\hline\hline
\textbf{0} & $0.3$, $1.1$ & 0.71 & $0.66\pm 0.03$ & $1.34\pm 0.08$ & $2.05\pm0.18$ \\
\hline
\textbf{2.2} & $1.1$ & 0.88 & $0.71\pm0.04$ & $1.12\pm 0.10$ & $1.81\pm 0.12$ \\ 
\textbf{3.0} & $0.7$ & 1.11 & $0.87\pm0.05$ & $1.05\pm 0.11$ &  $1.69\pm 0.11$\\ 
\hline
\textbf{7.5} & $1.1$& 1.44 & $0.95\pm0.13$ & $0.82\pm 0.09$ & $1.45\pm 0.14$ \\ 
\textbf{8.1} & $0.7$ & 1.35 & $0.94\pm0.04$ & $0.91\pm 0.12$ & $1.39\pm 0.06$ \\ 
\hline
\textbf{20} & $0.5$ & 1.50 & $1.17\pm0.06$ & $0.45\pm 0.13$ & $1.03\pm 0.32$\\
\hline \hline
\end{tabular}
\caption{The numerical coefficients found for fitting each of the cohesive grains to the empirical power laws from Lajeunesse \textit{et al.} \cite{2005_lajeunesse} as denoted in \eqref{eq:1}. Uncertainties display the variation of fits.}
\label{table:table2_LajScal}
\end{center}
\end{table}

\begin{table}
\begin{center}
\begin{tabular}{||m{6em}|m{6em}||m{1.75cm}|m{1.75cm}||m{1.75cm}|m{1.75cm}||m{1.75cm}|m{1.75cm}||} 
\hline\hline
\textbf{$\mathrm{Co}$} & $d$ (mm)& $\bar{\alpha}$&      A        &  $\bar{\beta}$ &       B        & $\bar{\gamma}$ & C \\ [0.6ex]
\hline\hline
\textbf{0}      & $0.3$, $1.1$  & $0.68\pm 0.03$ & $-0.05\pm0.04$ & $1.38\pm 0.08$ & $0.88\pm0.11$  & $2.15\pm 1.21$ & $0.48\pm 0.25$ \\
\hline
\textbf{2.2}    & $1.1$         & $0.76\pm 0.07$ & $-0.05\pm0.05$ & $1.08\pm 0.12$ & $1.05\pm 0.08$ & $1.69\pm 0.65$ & $0.54\pm 0.20$ \\ 
\textbf{3.0}    & $0.7$         & $0.93\pm 0.06$ & $-0.06\pm0.05$ & $1.19\pm 0.09$ & $0.78\pm 0.11$ & $1.43\pm 0.22$ & $0.59\pm 0.07$\\ 
\hline
\textbf{7.5}    & $1.1$         & $1.08\pm 0.41$ & $-0.13\pm0.02$ & $0.81\pm 0.14$ & $1.02\pm 0.24$ & $1.38\pm 0.68$ & $0.52\pm 0.24$ \\ 
\textbf{8.1}    & $0.7$         & $1.04\pm 0.10$ & $-0.08\pm0.05$ & $1.03\pm 0.10$ & $0.75\pm 0.15$ & $1.24\pm 0.27$ & $0.56\pm 0.09$ \\ 
\hline
\textbf{20}     & $0.5$         & $1.27\pm 0.17$ & $-0.06\pm0.05$ & $0.31\pm 0.30$ & $1.41\pm 1.40$  & $0.59\pm 0.20$ & $0.81\pm 0.19$ \\
\hline \hline
\end{tabular}
\caption{The numerical coefficients and {exponents} found for fitting each of the cohesive grains to the general power laws as denoted in (\ref{eq:1_general}). Uncertainties display the variation of fits.}
\label{table:table3_freeScal}
\end{center}
\end{table}

Table \ref{table:table2_LajScal} summarizes the results of the final deposit for an axisymmetric deposit of a cylindrical column for the range of tested cohesion. The critical aspect ratio between the two trends {given by Eq. \eqref{eq:1}} is denoted ${a_{c,\, H}}$ and {defined here} as the mean $a$ between the two trials corresponding to the largest $a$ for which $H_{\infty} = H_0$ and the smallest $a$ for which $H_{\infty} < H_0$. {Assuming} the power-laws for the cohesionless grains from Lajeunesse \textit{et al.} \cite{2005_lajeunesse} as denoted in \eqref{eq:1} {to remain valid in the cohesive case}, {the coefficients} $\alpha$, $\beta$, and $\gamma$ are {fitted} for each cohesion {tested}. Since {the influence of cohesion on} the critical aspect ratio for the runout trends{, $a_{c,\,\Delta}$,} is not as apparent {as for $a_{c,\, H}$}, ${a_{c,\,\Delta}} = 3$ is used following Lajeunesse \textit{et al.} \cite{2005_lajeunesse} for cohesionless grains.
\smallskip

This assumption that the effects of cohesion can be captured by modifying only the numerical prefactors is tested {here}. For instance, the numerical simulations of rectangular submerged cohesive collapses of Zhu \textit{et al.} \cite{2022_zhu} show changes in the power law {exponents} due to cohesion. At first glance, the radial runout for $\mathrm{Co}=20$ appears to follow a power law different from the other data{, as illustrated in Fig. \ref{fig:Fig6_FinalMorphology_3D}(b)}. Consequently, alternative fits are considered {here}. Instead of enforcing the {exponents} of the power laws, we also {presents} fits in the form:

{\allowdisplaybreaks
\begin{alignat}{4}
  \label{eq:1_general}
	\frac{H_\infty}{R_0} = & \left\{ \begin{array}{ll}
	\displaystyle a\\[8pt]
	\displaystyle \bar{\alpha}a^{\rm A}
	\end{array}\right.
    & \begin{array}{ll}
	\displaystyle \ \mathrm{for\ }  a \lesssim {a_{c, \, H}},\\[8pt]
	\displaystyle \ \mathrm{for\ }  a \gtrsim {a_{c, \, H}},
	\end{array}
    \qquad
	\frac{\Delta R_\infty}{R_0} = & \left\{
	\begin{array}{ll}
	\displaystyle \bar{\beta} a^{\rm B}\\[8pt]
	\displaystyle \bar{\gamma} a^{\rm C}
	\end{array}\right.
    & \begin{array}{ll}
	\displaystyle \ \mathrm{for\ }  a \lesssim {a_{c, \, \Delta}},\\[8pt]
	\displaystyle \ \mathrm{for\ }  a \gtrsim {a_{c, \, \Delta}},
	\end{array}
\end{alignat}
} where $\bar{\alpha}, \bar{\beta}, \bar{\gamma}$ are numerical coefficients and $\rm A, \, B, \, C$ are also allowed to change. Results for all our grains are shown in Table \ref{table:table3_freeScal}. In general, the powers proposed by Lajeunesse \textit{et al.} \cite{2005_lajeunesse} with A, B, and C equal to 0, 1, and 1/2, respectively, work well for our small and moderate cohesions. The case with $\mathrm{Co}=20$ deviates from these powers considerably. Since columns of this cohesion are stable for $a \lesssim 1.5$, only four points between $1.5 < a < 3$ are available to determine B, showing large consequent variation. The best fitting power law for $a > {a_{c, \Delta}}$ is using $\rm C \approx 4/5$. This suggests that at large cohesion, the power laws may show differences. This is expected as the material, when adequately cohesive, will likely go through brittle collapses, such as those considered in the numerical simulations by Langlois \textit{et al.} \cite{2015_langlois}.  
\smallskip

\begin{table}
\begin{center}
\begin{tabular}{||m{6em}|m{6em}||m{2cm}|m{2cm}||m{2cm}|m{2cm}||} 
\hline\hline
\textbf{$\mathrm{Co}$} & $d$ (mm) & ${a_{c,\, H}}$ & $\delta$ & $\varepsilon$ & $\zeta$ \\ [0.5ex]
\hline\hline
\textbf{0} & $0.3$, $1.1$ & 0.84 & 0.91$\pm$0.01 & 1.71$\pm$0.06 & 2.43$\pm$0.08 \\
\hline
\textbf{2.2} & $1.1$ & - & 0.93$\pm$0.04 & 1.76$\pm$0.09 & 2.42$\pm$0.04 \\ 
\hline
\textbf{7.5} & $1.1$ & 1.07 & 1.22$\pm$0.11 & 0.98$\pm$0.16 & 1.50$\pm$0.17 \\ 
\textbf{8.1} & $0.7$ & 1.06 & 1.03$\pm$0.06 & 1.16$\pm$0.03 & 1.62$\pm$0.12 \\ 
\hline
\textbf{20} & $0.5$ & 1.72 & 1.29$\pm$0.11 & 0.68$\pm$0.18 & 1.18$\pm$0.15 \\
\hline \hline
\end{tabular}
\caption{The numerical coefficients found for fitting each of the cohesive grains to the empirical power laws from Lajeunesse \textit{et al.} \cite{2005_lajeunesse} as denoted in Eq. \eqref{eq:2}. Uncertainties display the variation of fits.}
\label{table:table4_LajScal_2D}
\end{center}
\end{table}

\begin{table}
\begin{center}
\begin{tabular}{||m{6em}|m{6em}||m{1.75cm}|m{1.75cm}||m{1.75cm}|m{1.75cm}||m{1.75cm}|m{1.75cm}||} 
\hline\hline
\textbf{$\mathrm{Co}$} & $d$ (mm)& $\bar{\delta}$&      L        &  $\bar{\varepsilon}$ &       M        & $\bar{\zeta}$ & N \\ [0.6ex]
\hline\hline
\textbf{0}      & $0.3$, $1.1$  & $0.84\pm 0.03$ & $0.39\pm0.02$ & $1.77\pm 0.12$ & $0.97\pm0.09$  & $2.09\pm0.17$ & $0.74\pm 0.08$ \\
\hline
\textbf{2.2}    & $1.1$         & $0.99\pm 0.09$ & $0.29\pm0.06$ & $1.77\pm 0.05$ & $0.99\pm 0.31$ & $2.50\pm 0.21$ & $0.65\pm 0.04$ \\ 
\hline
\textbf{7.5}    & $1.1$         & $0.94\pm 0.15$ & $0.50\pm0.09$ & $0.83\pm 0.26$ & $1.26\pm 0.36$ & $1.16\pm 0.8$ & $0.80\pm 0.35$ \\ 
\textbf{8.1}    & $0.7$         & $1.17\pm 0.11$ & $0.26\pm0.06$ & $1.15\pm 0.10$ & $1.07\pm 0.10$ & $1.48\pm 0.65$ & $0.71\pm 0.22$ \\ 
\hline
\textbf{20}     & $0.5$         & $1.37\pm 0.56$ & $0.30\pm0.22$ & $0.51\pm 0.25$ & $1.37\pm 0.61$ & $0.98\pm 1.30$ & $0.76\pm 0.71$ \\
\hline \hline
\end{tabular}
\caption{The numerical coefficients and powers found for fitting each of the cohesive grains to the general power laws as denoted in (\ref{eq:2_general}). Uncertainties display the variation of fits.}
\label{table:table5_freeScal_2D}
\end{center}
\end{table}

Similarly, for the 2D channelized collapses, Table \ref{table:table4_LajScal_2D} summarizes the results for the range of tested cohesions, assuming the presence of cohesion only changes the coefficients of the power laws given by Eq. \eqref{eq:2}. Instead of enforcing powers, again general {exponents can be} attempted to find the best fit of the form:
{\allowdisplaybreaks
\begin{alignat}{4}
  \label{eq:2_general}
	\frac{H_\infty}{L_0} = & \left\{ \begin{array}{ll}
	\displaystyle a\\[8pt]
	\displaystyle \bar{\delta}a^{\rm L}
	\end{array}\right.
    & \begin{array}{ll}
	\displaystyle \ \mathrm{for\ }  a \lesssim {a_{c, \, H}},\\[8pt]
	\displaystyle \ \mathrm{for\ }  a \gtrsim {a_{c, \, H}},
	\end{array}
    \qquad
	\frac{\Delta L_\infty}{L_0} = & \left\{
	\begin{array}{ll}
	\displaystyle \bar{\varepsilon} a^{\rm M}\\[8pt]
	\displaystyle \bar{\zeta} a^{\rm N}
	\end{array}\right.
    & \begin{array}{ll}
	\displaystyle \ \mathrm{for\ }  a \lesssim {a_{c, \, \Delta}},\\[8pt]
	\displaystyle \ \mathrm{for\ }  a \gtrsim {a_{c, \, \Delta}},
	\end{array}
\end{alignat}
} where $\bar{\delta}$, $\bar{\varepsilon}$ and $\bar{\zeta}$ are numerical coefficients, and L, M, and N are {free exponents}. Results with these general power laws are tabulated in Table \ref{table:table5_freeScal_2D}. {It should be recalled that, for} the experiments with cohesionless grains of Lajeunesse \textit{et al.} \cite{2005_lajeunesse}, L, M, and N are 1/3, 1, and 2/3, respectively. As can be seen in the table, these powers are generally within the error bars for all the cohesive grains as well. Small variations are found with $\mathrm{Co}$= 7.5 and 20, for these grains correspond with the wet grains. Particularly in the large $a$ cases, the capillary bridges between the grains and the confining walls seem to {slightly} change the behavior of such deposits. Meanwhile, experiments with CCGM more closely resemble the power-law trends of the cohesionless grains since the coating is non-reactive with the PMMA enclosure. Ultimately, across the two geometries, the power laws describing the morphology of the final deposit for cohesionless grains also describe {fairly well} the range of cohesion considered in this study. {Therefore, this demonstrates that, at first order, cohesion mainly affects the numerical prefactors of these scalings, which justifies the assumption adopted in the main text.}

\bibliography{collapse_biblio}
\end{document}